\documentclass[journal=jacsat,manuscript=article,layout=twocolumn ]{achemso}

\usepackage{multicol}
\usepackage{url}
\usepackage{hyperref}
\usepackage{caption}
\usepackage{amsmath}
\usepackage[dvipsnames]{xcolor}
\usepackage{graphicx}
\usepackage{algorithm}
\usepackage{algpseudocode}

\usepackage{caption}
\usepackage{graphicx}
\usepackage{subcaption}
\usepackage{overpic}

\usepackage[version=3]{mhchem} 
\usepackage[dvipsnames]{xcolor}
\usepackage{amsmath}

\author{Xiaomi Guo}
\affiliation[Tsinghua University]{State Key Laboratory of Low Dimensional Quantum Physics and Department of Physics, Tsinghua University, Beijing, 100084, China}
\alsoaffiliation[Aalto University]{Department of Applied Physics, Aalto University, Espoo, Finland}
\author{Lincan Fang}
\affiliation[Aalto University]{Department of Applied Physics, Aalto University, Espoo, Finland}
\author{Yong Xu}
\affiliation[Tsinghua University]{State Key Laboratory of Low Dimensional Quantum Physics and Department of Physics, Tsinghua University, Beijing, 100084, China}
\alsoaffiliation[FSC, China]{Frontier Science Center for Quantum Information, Beijing 100084, China}
\alsoaffiliation[RIKEN, Japan]{RIKEN Center for Emergent Matter Science (CEMS), Wako, Saitama 351-0198, Japan}
\author{Wenhui Duan}
\affiliation[Tsinghua University]{State Key Laboratory of Low Dimensional Quantum Physics and Department of Physics, Tsinghua University, Beijing, 100084, China}
\alsoaffiliation[IAS, THU]{Institute for Advanced Study, Tsinghua University, Beijing 100084, China}
\alsoaffiliation[FSC, China]{Frontier Science Center for Quantum Information, Beijing 100084, China}
\author{Rinke Patrick}
\affiliation[Aalto University]{Department of Applied Physics, Aalto University, Espoo, Finland}
\author{Milica Todorovi\'{c}}
\email{milica.todorovic@aalto.fi}

\affiliation[University of Turku]{Department of Mechanical and Materials Engineering, University of Turku, FI-20014 Turku, Finland}
\author{Xi Chen}
\email{xi.6.chen@aalto.fi}
\affiliation[Aalto University]{Department of Applied Physics, Aalto University, Espoo, Finland}

\title[An \textsf{achemso} demo]
  {Molecular conformer search with low-energy latent space}


\begin{document}

\begin{tocentry}

\begin{center}
\includegraphics[width=1.0\textwidth]{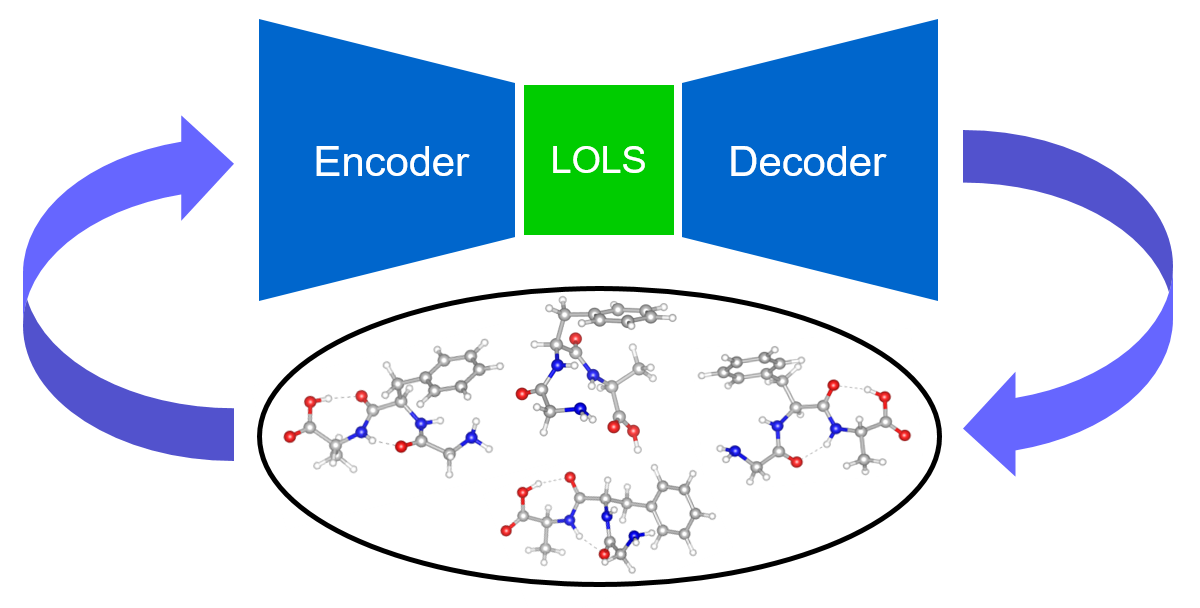}
\end{center}

\end{tocentry}

\begin{abstract}
  
Identifying low-energy conformers with quantum mechanical accuracy for molecules with many degrees of freedom is challenging. In this work, we use the molecular dihedral angles as features and explore the possibility of performing molecular conformer search in a latent space with a generative model named variational auto-encoder (VAE). We bias the VAE towards low-energy molecular configurations to generate more informative data. In this way, we can effectively build a reliable energy model for the low-energy potential energy surface. After the energy model has been built, we extract local-minimum conformations and refine them with structure optimization.  We have tested and benchmarked our low-energy latent-space (LOLS) structure search method  on organic molecules with $5-9$ searching dimensions. Our results agree with previous studies.

\end{abstract}

\section{Introduction}

Organic molecules are typically very flexible, and any molecule with rotatable bonds can adopt multiple energetically accessible conformations, each associated with different chemical and electronic properties. \cite{confer1, ES} Identifying the low-energy molecular conformers and determining their energy ranking is therefore a topic of great importance in computational chemistry\cite{CompChem}, cheminformatics \cite{cheminfor}, computational drug design \cite{confer-drug}, and structure-based virtual screening \cite{Screening}.  However, the dimension of configurational spaces and the complexity of energy landscapes increases drastically with the size of the molecule. This makes molecular conformer search one of the persistent challenges in molecular modeling. \cite{confer1,confer2}

A variety of methods and tools have been developed for molecular conformer search. Systematic methods use a grid to sample all possible torsion angles in a molecule. This approach is deterministic but limited to small molecules due to its poor scaling with increasing search dimensions. Conversely, methods such as Monte Carlo annealing\cite{MC}, minima hopping \cite{2004minima}, basin hopping \cite{BH2} and genetic algorithms \cite{GA} sample configurational space stochastically. Stochastic methods can be applied to larger molecules with high-dimensional search spaces, but due to the random nature of the process, extensive sampling is required to achieve convergent results. To balance the accuracy and computational cost, hierarchical methods which first scan a large portion of configurational space, and then refine the promising candidate with more costly and accurate computations have been developed
\cite{hiera1,hiera2}. Since simulation methods at different levels of accuracy may predict different potential energy surfaces (PES), a large number of structures still needs to be optimized at the higher level to avoid missing the true low-energy conformers. \cite{hiera1}

In recent years,  machine learning techniques such as artificial neural networks \cite{ML-M3,ML-NN}, Gaussian process regression (GPR) \cite{GRP2, GRP3, GP-PRB,ML-N1}, and machine-learned force fields \cite{ML-FF-AK} have been successfully applied to accelerate structure-to-energy predictions and geometry optimization for molecules. However, most of these schemes require training on large data sets, usually costly to compute with \textit{ab initio} methods.

In our recent work, we presented a new approach based on Bayesian Optimization and quantum chemistry methods for molecular conformer identification and ranking. \cite{Fang} We first  kept all bond lengths and angles fixed, and selected the dihedral angles as the features to form the search space. Then we employed the BOSS code \cite{BOSS, BOSS-web}  to actively learn the PES of the molecule by Bayesian Optimization iterative data sampling. After the PES converged, we analyzed the PES to extract the local minima locations and related structures, and optimized the structures with density funcational theory (DFT) and other post-processings. 
We have tested our method on cysteine, serine, tryptophan, and aspartic acid. The method shows both high accuracy and efficiency, and can be easily automated for extensive searches.
The excellent efficiency is partly due to learning the PES in the reduced conformational space of dihedral angles and only refining the local minima structures with DFT, and partly because Bayesian Optimization creates small and compact data sets. However, our method is not directly transferable to molecules with high-dimensional search spaces. The data required for building reliable PESs increases rapidly with search dimensions. With increasing data set size, the cost to compute the necessary data with quantum mechanical methods and to build the surrogate model of the PES in BOSS grows and eventually becomes prohibitively expensive.

\begin{figure*}[htpb]
    \centering
    \includegraphics[width=1.0\textwidth]{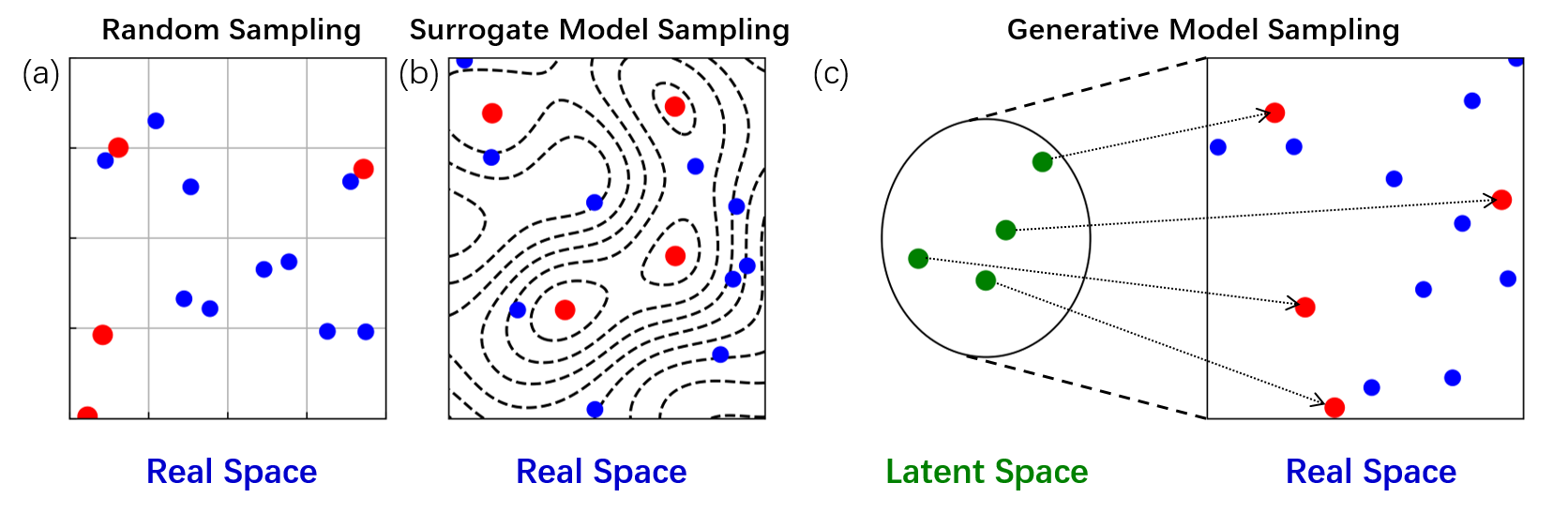}
    \caption{A viewer of sampling methods. The blue and red dots represent the acquired samples and candidates for next sampling. The dash lines in (b) represents the contour lines of the surrogate model. The green dots in (c) represent samples in latent space. The generator can map them to real space.}
    \label{figure:methods}
\end{figure*}

To address this challenge, we will explore the possibility of using a generative model to acquire samples in a latent space for molecular conformer search. We decided on variational auto-encoders (VAEs) as the generative model, because the neural network structure of VAEs is typically simple; and VAEs are equipped with a regularization term in the loss function to prevent over-fitting. VAEs combine an encoding neural network  (encoder) with a  decoding neural network  (decoder). The encoder compresses data from real space (here the space of dihedral angles) into a latent space. This compression ideally retains the essential data correlations  in  the  reduced  representation. The decoder maps latent vectors back to  the original representation.
Figure ~\ref{figure:methods} illustrates how sampling in latent space with a generative model (c) differs from conventional random sampling in real space (a) and from our previous approach of employing a surrogate model and an acquisition strategy (b). 

To sample more efficiently with our generative approach, we are steering the VAE towards low-energy molecular configurations during the training. The latent space then predominantly encodes information on the relevant, low-energy region of the PES. As in previous work, we use dihedral angles to represent the different molecular conformations. We also extract local minima structures and apply structure optimization only after a meaningful PES has been learned.

\begin{figure*}[htpb]
  \centering
  \includegraphics[width=1.0\textwidth]{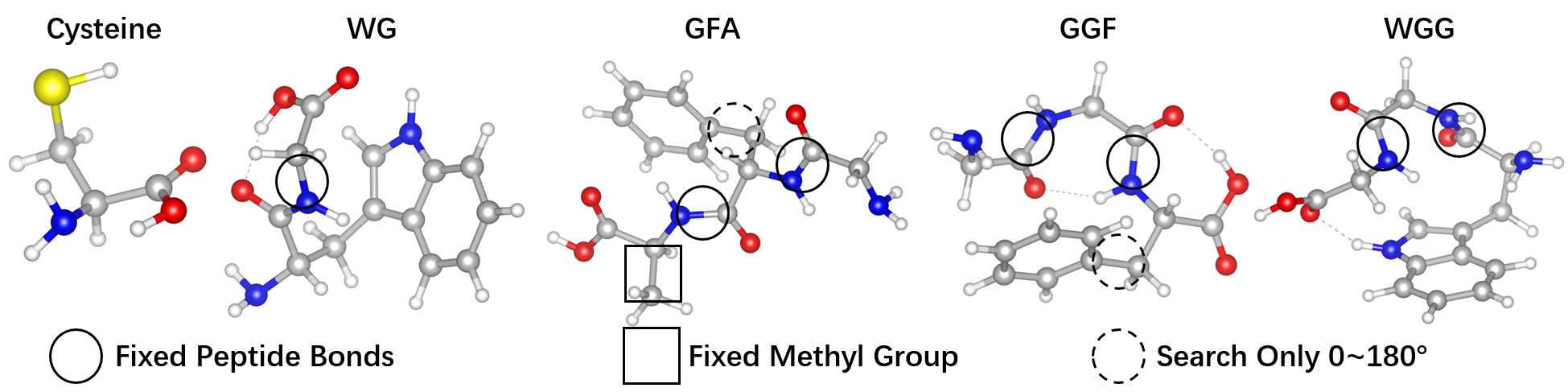}
  \caption{Ball-and-stick models of cysteine, tryptophyl-glycyl (WG), glycyl-phenylalanyl-alanyl (GFA), glycyl-glycyl-phenylalanyl (GGF) and tryptophyl-glycyl-glycyl (WGG). Red atoms denote oxygen, white hydrogen, gray carbon, blue nitrogen, and yellow sulfur. The dashed circle mark the dihedral angles that have a reduced search range of  [0$^\circ$, 180$^\circ$]. The solid circles and squares mark peptide bonds and dihedral angles that are kept fixed during sampling. All other dihedral angles belong to our space with their full range [0$^\circ$,$360^\circ$].}
  \label{picture:materials}
\end{figure*}

In brief, in this work we designed a low-energy latent-space (LOLS) structure search method for molecular comformer search and determined appropriate settings and suitable hyperparameters for it.  We tested LOLS on cysteine and four peptides tryptophyl-glycyl (WG), glycyl-phenylalanyl-alanyl (GFA), glycyl-glycyl-phenylalanyl (GGF) and tryptophyl-glycyl-glycyl (WGG) (Figure~\ref{picture:materials}).  The main reasons for choosing these molecules are: First, amino acids and peptides are important biomolecules. Second, peptides are very flexible and exhibit complex PESs, making them a challenging system for conformer search. Third, previous studies provide reference data. \cite{Fang, Exp, pep-database}. Another objective of our work is to gain insight into the nature and properties of latent space. For this, we visualize and analyze the latent spaces of cysteine and GFA. Our method and our results will be presented in the following sections.

\section{Methods}

\begin{figure*}[htpb]
    \centering
    \includegraphics[width=0.8\textwidth]{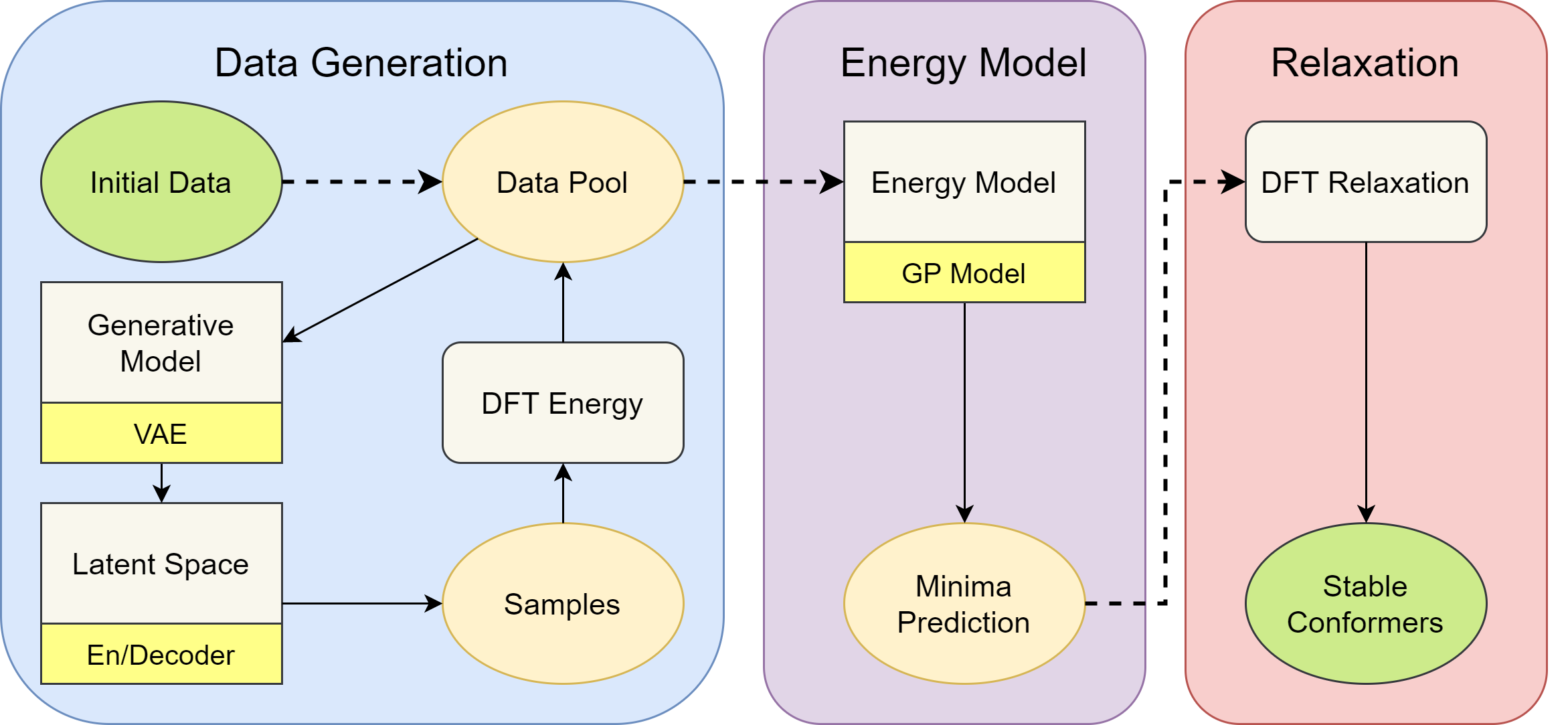}
    \caption{The LOLS workflow starts from initial data and finishes with the structures and energies of stable conformers. The eclipses represent data, the rectangles machine learning models, and the rectangles with round corners represent DFT calculations.}
    \label{figure:abstractworkflow}
\end{figure*}

Our LOLS method consists of three steps (Figure~\ref{figure:abstractworkflow}). In  step 1, we employ an active learning approach to generate data on-the-fly. 
We combine two strategies to steer the generative model towards generating more low-energy data, which helps us build a compact and reliable model for the low-energy regions of the PES. Strategy one is data processing. We scale the energy of training data with a non-linear function and exclude high-energy data. Strategy two attributes more weight to lower energy data in the loss function of the generative model. Both strategies will be discussed in the following sections. In step 2, we build a Gaussian process (GP) regression model in real space. We extract the local minima from the GP and use them to initialize DFT geometry optimizations. In step 3, the candidate structures are further optimized with DFT structure relaxation. Details of our method will be explained in the following sections.

\subsection{Data Generation Loop}

The left part of Figure~\ref{figure:abstractworkflow} shows the data generative loop we designed for sampling informative data. The ``data pool" is initialized with an initial data set, in which each data point represents the dihedral angles and DFT energy of a conformation. Then we set up an  active learning approach and iteratively acquire samples from the latent space. For each new sample, the structural features are decoded by the VAE into real space, then the energy is calculated with DFT. As we add new samples to the data pool, we keep retraining the VAE.

Each time we carry out three parallel runs to average out the effects of randomization in the sampling method, and continue the data generation loop up to a preset maximum number of iterations. If the global minimum and at least
70\% of the reference targets are found, we stop
the data generation, otherwise we continue.
The details are explained in sections  ``VAE and latent space" and ``sampling method".

\begin{figure*}[htpb]
    \centering
    \includegraphics[width=0.9\textwidth]{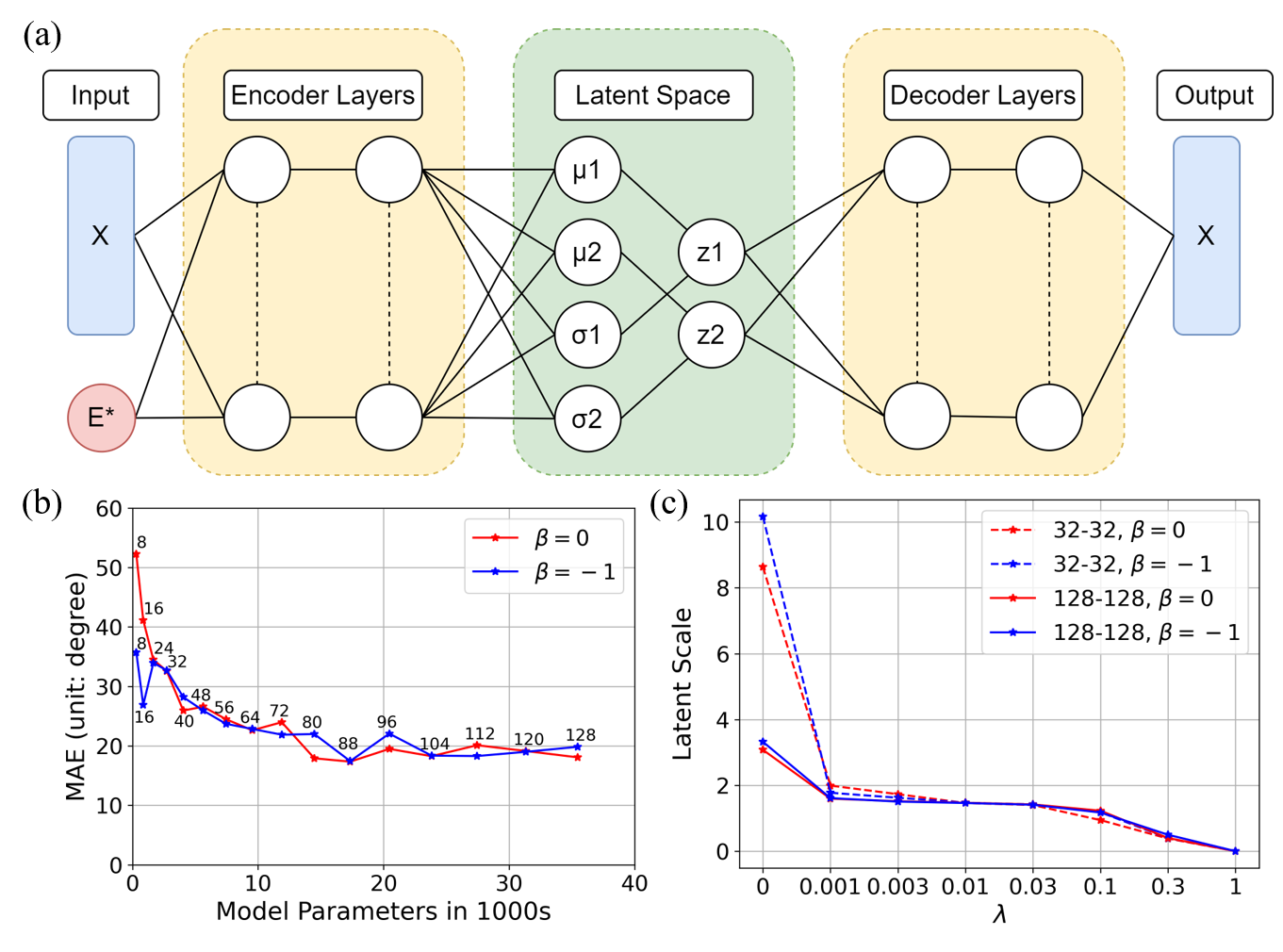}
    \caption{(a) Architecture of our variational auto-encoder. The input X consists of molecular dihedral angles, and input $E^*$ is the scaled energy of the corresponding structure. The output X are again dihedral angles, but no energy. Both encoder and decoder have two layers with the same layer size. The mean ($\mu$) and variance ($\sigma^2$)  are the outputs of the encoder and the normal distribution is taken for samples $z$. (b) The mean absolute difference (MAE)  between input and output dihedral angles for the VAE models with different numbers of trainable parameters.  The text near the data points shows the numbers of neurons of each fully connected layer in the en/decoder ($layersize$). During the test of $layersize$, $\lambda$ is fixed to 0.01.  (c) The relationshop between the latent-space scale $L$ and the hyperparameter $\lambda$. The test were performed with 32-32, 128-128 neural networks and $\beta = 0$, $-1$.
    \label{figure:vae_setting}}
\end{figure*}

\subsubsection{VAE and latent space}

Figure~\ref{figure:vae_setting}(a) shows the architecture of our VAE. The encoder layers reduce the dimension of the input data and cast the input data into a distribution in latent space with mean $\mu_j$ and variance $\sigma^2_j$ ($j$ represents the axis number of latent space).  During the training stage, the vector $z$ in latent space is generated by $z_{j}=\mathcal{N}(\mu_{j}, \sigma^2_{j})$,\cite{kingma2014autoencoding} where $\mathcal{N}$ is the normal distribution. The vector $z$ can be mapped back to real space by the decoder layers.  

\textbf{Data preprocessing.}  The raw data includes the dihedral angles of sampled molecular structures and their DFT-calculated energy $E$. We preprocess the data in two stages. In stage one, the dihedral angles are normalized from [0,360] to [-1, 1], and the total DFT energy is scaled according to the following equation

\begin{equation}
\begin{array}{l} 
  E^*=\left\{\begin{matrix}
  E-E_0 & E \le  E_0\\ 
  \ln(1 + E - E_0) & E > E_0
\end{matrix}\right.
\end{array}
\label{eq:energy}
\end{equation}
where $E_0$ is a threshold energy that is used to shift the DFT energies close to zero. $E_0$ is system dependent but once chosen is kept constant for the same molecule (see Table~\ref{table:diff}). We adopt the logarithmic function in Equation~\ref{eq:energy} to scale down high energies ($E>E_0$), because we are primarily interested in the low energy region and wish to avoid high energy regions that can obstruct model fitting.

In stage two, data with a scaled energy larger than $E^*_{\mathrm{max}} = \mathbf{mean}(E^*) + \alpha\times\mathbf{std}(E^*)$ is excluded from the training set of the VAE, since the corresponding structures frequently exhibit steric clashes and are therefore not relevant. In this work, we set the cutoff threshold $\alpha$ = 2, which resulted in a data exclusion of $3-6\%$ from the training set of the VAE. The excluded data is usually  5 to 25 eV above the global minimum, and was still kept in the data pool and used to build the energy model in step 2. 

\textbf{Loss function.} The trainable parameters of the VAE are optimized by minimizing the total loss function, which consists of two contributions 

\begin{equation}
\begin{aligned} 
\delta_{\mathrm{total}}=\delta_{\mathrm{rec}}+\lambda\delta_{\mathrm{reg}}.
\end{aligned}
\end{equation}
The first part is the reconstruction loss ($\delta_{\mathrm{rec}}$), which forces the encoder-decoder pair to minimize information loss (i.e. minimize the difference between input and output). The second part is the  regularization  ($\delta_{\mathrm{reg}}$) that confines the latent space by forcing the encoder output towards a standard normal distribution. $\lambda$ is a hyperparameter that controls the ratio between the two loss terms.

To make the VAE more sensitive to low-energy structures, we weight the reconstruction loss term ($\delta_{\mathrm{rec}}$) with the corresponding scaled energy $\exp(\beta E^*)$,

\begin{equation}
\begin{aligned}
\delta_{\mathrm{rec}} =
\frac{\sum_{i=1}^{N} {\exp(\beta E_i^*)\times \mathbf{Diff}(x_i^\mathrm{in}, x_i^\mathrm{out})}}{\sum_{i=1}^{N} {\exp(\beta E_i^*)}}
\end{aligned}
\label{eq:ReLoss}
\end{equation}
where $\beta$ is a hyperparameter which will be explored and discussed later. In this work, we varied $\beta$ from 0 to -3. $i$ refers to the $i$th training data and $N$ is the size of the training data. $\mathbf{Diff}(x_i^\mathrm{in}, x_i^\mathrm{out})$ returns the difference between input and output for the $i$th training data. Since our VAE does not output the scaled energy, we define  $\mathbf{Diff}$ only in terms of the scaled dihedral angles 
\begin{equation}
\begin{aligned}
\mathbf{D}&\mathbf{iff}(x^\mathrm{in}_{i}, x^\mathrm{out}_{i}) =\\&\frac{1}{D}  \sum_{j=1}^{D}(((x_{ij}^\mathrm{in}-x_{ij}^\mathrm{out}+1)\mod2)-1)^2.
\label{eq:MSE}
\end{aligned}
\end{equation}
Here $D$ refers to the number of dihedral angles and $j$ to the $j$th input and output vectors.

The regularization term ($\delta_{\mathrm{reg}}$) can be expressed as the Kulback-Leibler (KL) divergence ($\delta_{\mathrm{kld}}$) between the returned distribution and a standard Gaussian \cite{kingma2014autoencoding}. 
According to  Ref.  \citenum{kingma2014autoencoding}, the KL divergence is calculated by the encoder output mean $\mu_{ij}$ and variance $\sigma^2_{ij}$, where $i$ is the $i$th training data, $j$ the axis number of latent space and $d$ is the dimension of latent space
\begin{equation}
\begin{aligned}
\delta_{\mathrm{kld}}=
\frac{1}{Nd} \sum_{i=1}^{N} \sum_{j=1}^{d} -\frac{1}{2}(1+\log \sigma_{ij}^2 - \mu_{ij}^2 - \sigma_{ij}^2).
\end{aligned}
\label{eq:KLD}
\end{equation}
The total loss function ($\delta_{\mathrm{total}}$) in our work is
\begin{equation}
\begin{aligned} 
\delta_{\mathrm{total}}=\delta_{\mathrm{rec}}+\lambda\delta_{\mathrm{kld}}.
\end{aligned}
\label{eq:total} 
\end{equation}

Next, we will select a suitable value for $\lambda$ and the right neural network settings for the cysteine data set we generated in our previous work \cite{Fang}.
The data set consists of 800 cysteine structures and their corresponding DFT energies from a BOSS run. We refer to this data set as CYS800. The dihedral angle and energy distributions of this data set are shown in Figure S1. 

\textbf{Neural network configurations.} We chose 2 as the latent space dimension, for the simple reason that two dimensions are convenient to visualize. Visualizing and analyzing the latent space will help us gain insight into the nature of the latent space and develop suitable sampling methods. It remains an open question if  increasing the dimension of latent space would help sample more informative data  and thus increase the efficiency of the approach. We will return to this question in future work.

For both encoder and decoder, we used two fully connected layers of the same size and ReLU as activation function. We varied the number of  neurons in each fully connected layer in the encoder or decoder ($layersize$) from 8 to 128 and checked the mean absolute error (MAE) between inputs and outputs. The  CYS800 data set was used in all the tests. Similar to Equation~\ref{eq:MSE}, the MAE is defined as
\begin{equation}
\begin{aligned}
\mathbf{MA}&\mathbf{E}(x^\mathrm{in}_{i}, x^\mathrm{out}_{i}) =\\
&\frac{1}{D}  \sum_{j=1}^{D} \left | ((x_{ij}^\mathrm{in}-x_{ij}^\mathrm{out}+1)\mod2)-1 \right |.
\label{eq:MAE}
\end{aligned}
\end{equation}

In Figure~\ref{figure:vae_setting}(b) we show the MAE as a function of the number of neural network parameters, which is determined by the $layersize$. The MAE decreases with increasing $layersize$, but eventually converges around $20^{\circ}$. We believe that with a higher dimensional latent space (i.e., less information loss) we could further reduce the MAE, but we deemed $20^{\circ}$ sufficient for our purposes. We therefore picked a $layersize$ of 80 for cysteine and extended it to 128 for other molecules in this work with higher search dimensions.

The VAE was trained for 100,000 epochs to ensure the convergence of the total loss function $\delta_\mathrm{total}$ (Figure S2). The value of the energy weight hyperparameter ($\beta=0$ or $\beta=-1$) has no significant effect on the MAE for the CYS800 data set, as shown in Figure~\ref{figure:vae_setting}(b). However, $\beta$ will play an important role in the active learning workflow (shown in Figure~\ref{figure:abstractworkflow}). We will discuss its effect in the results part.

\textbf{Loss ratio $\lambda$ and latent space.} After the training is finished, the encoder maps the training data into the latent space as the latent-space data $z_{ij} = \mu_{ij}$. The encoder output variances $\sigma^2_{ij}$ are only used in the reparameterization during the training stage and  ignored after training.  The hyperparameter $\lambda$ controls the ratio between the reconstruction loss and the KL-divergence, thus determining the shape and distribution of the latent-space data. We introduce the latent-space scale $L$ to measure the size of the latent space
\begin{equation}
L=\sqrt[]{\frac{1}{N}  \sum_{i=1}^{N}\sum_{j=1}^{d} {\mu_{ij}^2}}\text{ .}
\end{equation} 
Figure~\ref{figure:vae_setting}(c) shows that $L$ varies by one order of magnitude for $\lambda$ between 0 and 1. Between $\lambda=$ 0.001 and 0.03, $L$ stabilizes around 1.47 and changes little,  indicating we should pick $\lambda$ from this region. In this range, $L$ is also almost independent of the size of the neural network. 

Figure S3 shows the data distribution in latent space for different $\lambda$ values. The shape and size of latent spaces are highly dependent on $\lambda$. When $\lambda=0.01$, the latent-space data distributes uniformly inside a circle (Figure~\ref{figure:sample}), which may benefit sampling.   Therefore, we set $\lambda = 0.01$ for all networks in the following.  

\subsubsection{Sampling method}

After generating the latent space, we can sample it. Every sample will be decoded into dihedral angles to reconstruct the atomic structure in real space. Then the DFT energy of this structure is calculated. The combination of scaled dihedral angles and DFT energy $(x,E^*)$ is collected as new data. 

\begin{figure}[htbp]
    \centering
    \includegraphics[width=0.4\textwidth]{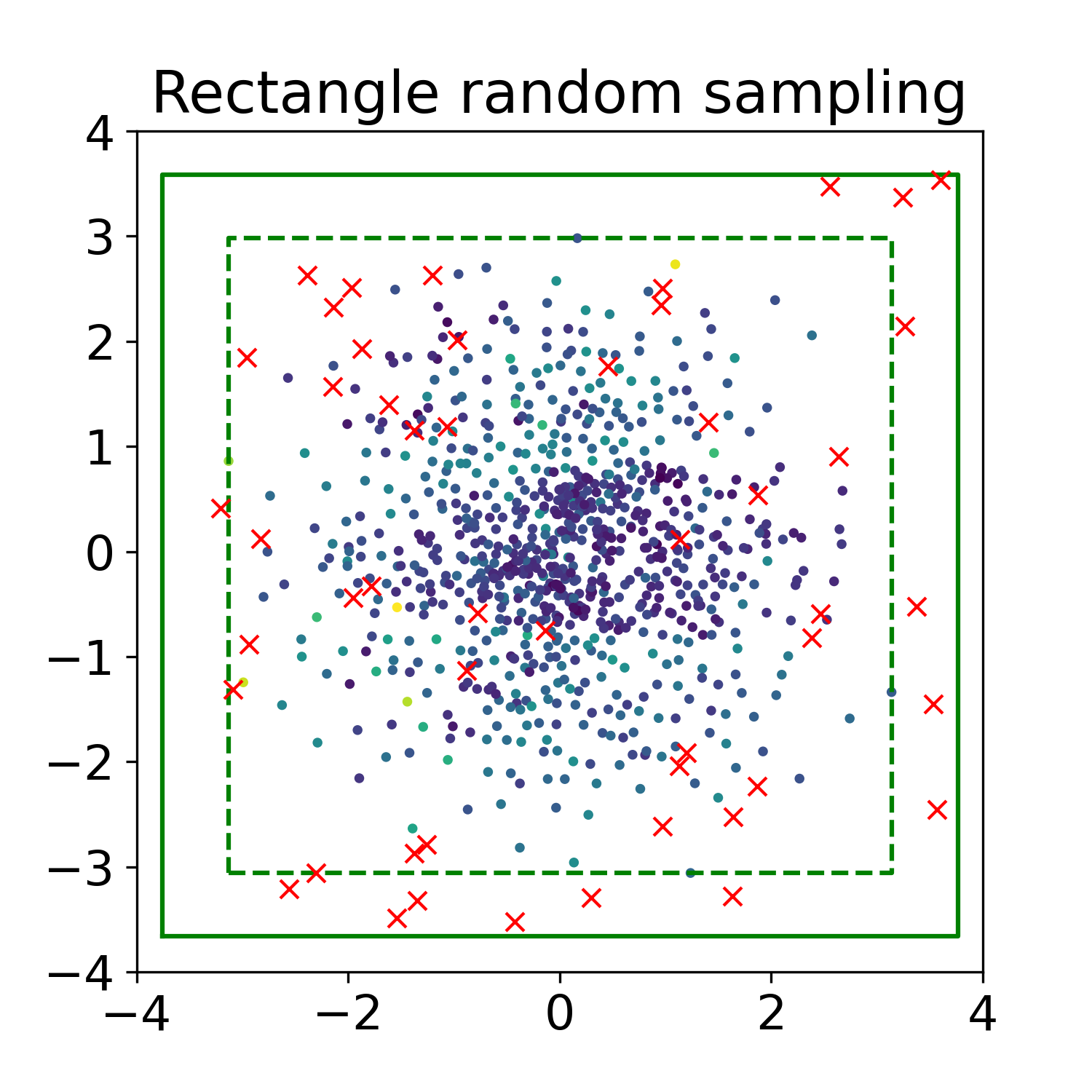}
    \caption{ The dashed and solid lines show the minimal rectangle and the extended rectangle, and the red crosses show a new sampling batch inside the extended rectangle. The latent space is the same as in Figure S3.}
    \label{figure:sample}
\end{figure}

We use a random sampling method to pick new structures from latent space. We had considered building a surrogate model of latent space with BOSS and sampling from its acquisition function, but the complex structure of latent space (which will be discussed in more detail in the ``Results and Discussion" section) does not lend itself to more advanced sampling methods. More specifically, we use a rectangle random sampling method (Figure~\ref{figure:sample}), which contains the following steps. First, we create a minimal rectangle that covers all of the latent-space data. Then we increase the width and height of the minimal rectangle with an expansion rate. The expansion rate is a hyperparameter that can be varied. We use a rate of 20\% in this work, which balances sampling from known latent space areas with the need to explore unknown areas away from available latent-space data. 
Finally, we choose positions randomly in the extended rectangle as samples.

In LOLS, the generation loop will keep running until the number of iterations reaches the preset maximum. At each iteration, the VAE is retrained and a data batch is acquired. These newly acquired data points are added to the data pool for training the new VAE in the next iteration. In this work, we fix the batch size in each iteration to 50, which is small enough to track changes in latent space and large enough to effect a change in the VAE.

\subsection{Energy Model}

We fit a surrogate model in real space after every $k$ iterations of the generation loop.
We call this the ``energy model'' as it establishes a relation between the dihedral angles and the energy.  $k$ is the energy model interval.  Here we choose $k$ = 5 for cysteine and $k$ = 20 for other molecules, which helped us find the relevant conformers without performing too many structure optimizations. The number of optimized structures is about 10-15\% of the number of samples  (See Table~\ref{table:result}). We could use a smaller $k$ to build more energy models and extract more local minima, but this would also require performing more DFT structure optimizations in step 3.

We use BOSS \cite{BOSS-web} to fit a GP to the energy model. The kernel is set to  standard periodic (STDP) to account for the periodicity of the dihedral angles, with inverse gamma priors employed to stabilise kernel hyperparameters. The $noise$ is set to 0.001 eV, comparable to the accuracy of DFT calculations. We set an uninformative prior on the GP mean to avoid biasing the model. After the energy model in real space is built by BOSS, we take the training data as the initial positions and apply the conjugate gradient method to find local minima. Only  different local minima are kept and  duplicates are purged. In accordance with our previous work, we fully optimize all molecular degrees of freedom with DFT for only these unique minima structures.

\subsection{DFT Method}
In this work, we employed the all-electron code FHI-aims  \cite{FHI-AIMS,Aims-efficientLC,AIMS-HYBcl} for all DFT calculations. We used ``tight" numerical settings, ``tier 2" basis sets, the PBE exchange-correlation functional \cite{PBE} and many-body dispersion (MBD) van de Waals corrections \cite{MBD}. For a few structures, in which two or more atoms come too close to each other, the FHI-aims single-point calculations fail. We consider these structures invalid (steric clashes). For different molecules, $3-6\%$ of samples were invalid and we omitted them.

For geometry optimization, a geometry was considered to be converged when the maximum residual force was below 0.01 eV/{\AA}. We stopped geometry optimization after a maximum of 200 steps to reduce the calculation costs. Any structure that is not converged after 200 steps is excluded.  For cysteine, all structures are converged in less than 200 relaxation steps, but for larger molecules, $5-20\%$ of structures do not converge (see Table~\ref{table:result}). 

\subsection{Complete Workflow}
Algorithm 1 shows the complete workflow of LOLS. We have defined the parameters $initdata$, $layersize$, $E_{0}$, $\alpha$, $\beta$, $\lambda$, $k$, and the $noise$ in the previous sections. In addition, $M$ represents the maximum iterations. 
\begin{algorithm}[htp]
\caption{Complete workflow}
\begin{algorithmic}[1]
\Require $initdata, layersize, E_{0}, \alpha, \beta, \lambda,$
$noise, M, k$
\State $\textbf{DataPool} = initdata$
\State $\textbf{StableComformers} = \emptyset$
\For {$i = 1 \ldots M$}
    \State $\textbf{Initialize}(\textbf{VAE}, layersize)$
    \State $data = \textbf{Trim}(\textbf{DataPool}, E_{0},\alpha)$
    \State $\textbf{Optimize}(\textbf{VAE}, \beta, \lambda, data)$
    \State $latent = \textbf{VAE} \rightarrow Encode(data)$
    \State $sample = \textbf{TakeSamples}(latent)$
    \State $decoded = \textbf{VAE} \rightarrow Decode(sample)$
    \For {$vector \in decoded$}
        \State $atoms = \textbf{Vec2Atoms}(vector)$
        \State $energy = \textbf{DFTEnergy}(atoms)$
        \State $\textbf{DataPool} \leftarrow \{vector, energy\}$
    \EndFor
    \If {$(i\equiv 0\mod{k})$}
        \State $\textbf{Initialize}(\textbf{GP}, \textbf{DataPool})$
        \State $\textbf{Optimize}(\textbf{GP}, noise)$
        \For {$vector, energy \in \textbf{DataPool}$}
            \State $\textbf{Optimize}(vector, \textbf{GP})$
            \State $atoms = \textbf{Vec2Atoms}(vector)$
            \State $\textbf{Optimize}(atoms)$
            \State $\textbf{StableComformers} \leftarrow atoms$
        \EndFor
    \EndIf
\EndFor
\State \textbf{Return StableComformers}
\end{algorithmic}
\end{algorithm}

We applied our LOLS method to cysteine and the peptides WG, GFA, GGF and WGG. Figure~\ref{picture:materials} shows how we chose the dihedral angles as features.  The dihedral angles of the peptide bonds in WG, GFA, GGF and WGG are fixed at $180^{\circ}$ for the trans conformation
because they usually have lower energy than the cis isomers. For GFA and GGF, the dihedral angles of the benzene rotation are only searched from $0^{\circ}$ to $180^{\circ}$ due to symmetry. For GFA, the dihedral angle of the methyl rotation is fixed at $180^{\circ}$. The final dimension of features for cysteine, WG, GFA, GGF, and WGG are 5, 7, 9, 9 and 9.  

The parameters in Table~\ref{table:share} are shared by all molecules. We do not fine-tune them for individual molecules because all the molecules in this work are small and organic. The molecule-dependent parameters are shown in Table~\ref{table:diff}. 

We could initialize LOLS with random data. However, since BOSS performs active learning for optimal knowledge gain and BOSS sampling is very fast for small amounts of data, we use samples from one BOSS run as the initial data in this work. The initial data size is also shown in Table~\ref{table:diff}. 

During testing on cysteine, we noticed that some targets that were correctly identified at a certain point would disappear, if we continued iterating (See Figure S4), due to statistical fluctuations of GP fitting. Because of this observation, we not only take the result from the final energy model with maximum data size but also from previous energy models.

\begin{table}[htp]

\begin{tabular}{|l|l|l|}
\hline
                    & name                    & value   \\ \hline
VAE      & latent dimension        & 2       \\ \cline{2-3} 
                    & cutoff threshold ($\alpha$)    & 2       \\ \cline{2-3} 
                    & energy weight ($\beta$) & 0,$-1$,$-3$ \\ \cline{2-3} 
                    & loss ratio ($\lambda$)  & 0.01    \\ \cline{2-3} 
                    & training epochs         & 100     \\ \hline
Sampling & expansion rate          & 20\%    \\ \cline{2-3} 
                    & batch size         & 50      \\ \hline
GP model & kernel                  & STDP    \\ \cline{2-3} 
                    & fitting noise           & 0.001   \\ \hline
\end{tabular}

\caption{General parameters of LOLS used for all the molecules in this work}
\label{table:share}
\end{table}

\begin{table*}[]
\begin{tabular}{|l|c|c|c|c|c|}
\hline
name                          & Cysteine & WG  & GFA    & GGF    & WGG    \\ \hline
search dimensionality            & 5        & 7   & 9      & 9      & 9      \\ \hline
initial data size             & 100      & 350 & 350    & 350    & 350    \\ \hline
en/decoder layer size  ($layersize$)       & 80       & 128 & 128    & 128    & 128    \\ \hline
maximum iteration ($M$)           & 40       & 120  & 120    & 120    & 140    \\ \hline
energy model interval ($k$) & 5        & 20  & 20     & 20     & 20     \\ \hline
threshold energy ($E_0$/eV) & -19635    & -24320   &  -27467 & -26399  & -29977  \\ \hline
\end{tabular}
\caption{Molecule-dependent parameters of LOLS}
\label{table:diff}
\end{table*}

\section{Results and Discussion}

We applied LOLS to cysteine, WG, GFA, GGF and WGG. For cysteine, we mainly compared the results to our previous study \cite{Fang}, which used BOSS and quantum chemistry methods. The conformer structures in Ref. \citenum{Fang} obtained with the same DFT settings as this work were selected as targets for cysteine.
For the other molecules, we compared our results to the database generated by Valders \textit{et al.} 
\cite{pep-database}. The authors first ran molecular dynamics/quenching (MD/Q) simulations with  tight-binding DFT to scan the free energy surfaces and then recalculated the low-energy structures with high-level quantum chemistry methods. We reoptimized their structures in the database with our DFT functional and settings before using them as targets. The mean difference in dihedral angles between our reoptimzied and the geometries in Ref.~\citenum{pep-database} are generally less than $5^\circ$, except WG 03 ($22.6^\circ$), GGF 05 ($12.3^\circ$), GGF 13 ($11.9^\circ$), and WG 11 ($7.0^\circ$). 
Two structures are considered similar when the mean difference in the dihedral angles is less than 15$^\circ$. In the series of similar structures, only the structure with the lowest energy is kept. If the maximal difference of dihedral angles between one target and one of our results is less than 15$^\circ$, we state that the target has been reached. Otherwise, we consider that a new structure has been found.

\subsubsection{Cysteine}

\begin{figure*}[htpb]
  \centering
    \includegraphics[width=1.0\textwidth]{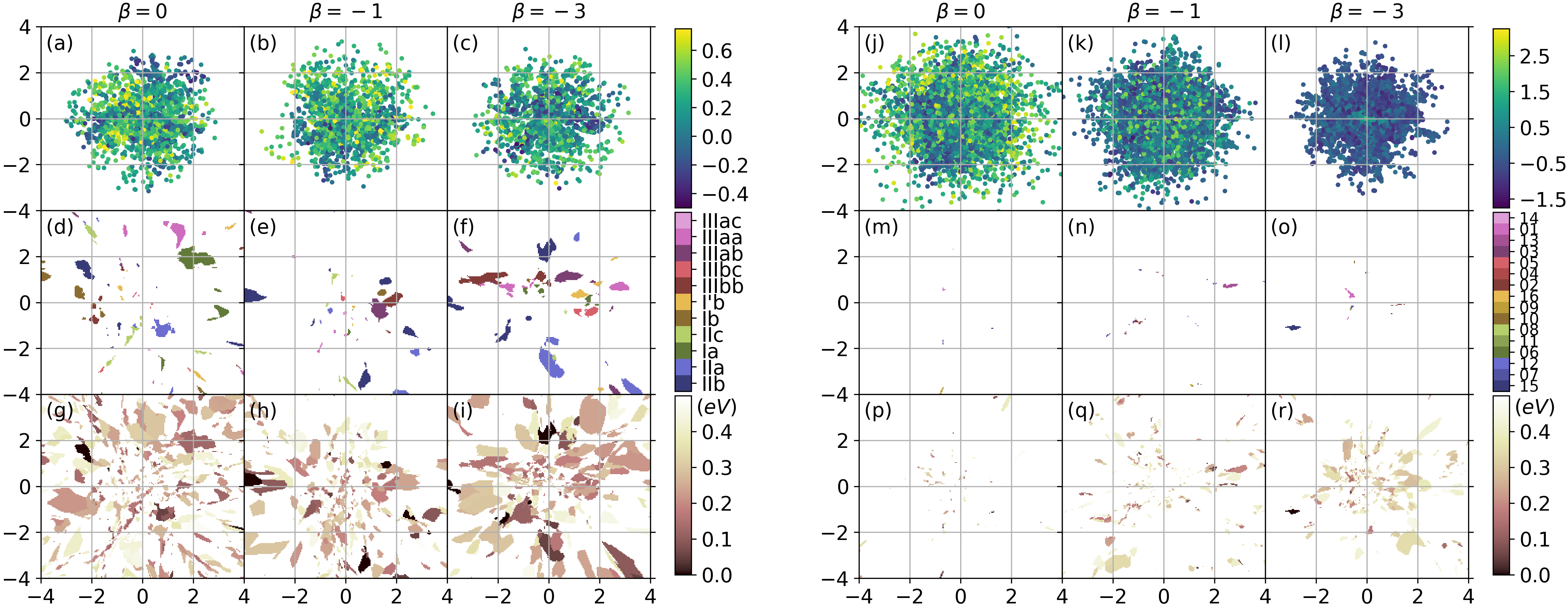}
    \caption{ The latent spaces of cysteine (left, (a)-(i)) and GFA (right, (j)-(r)) for different $\beta$ are visualized in three ways. In (a)-(c) and (j)-(l)
  the latent spaces are formed by the latent-space data generated by the encoders with different $\beta$. The color represents the scaled energy. In (d)-(i) and (m)-(r) the points in  latent space are decoded into real space, and the reconstructed structures are compared with a series of targets. If the mean difference of dihedral angles (MAE) between the reconstructed structure and the nearest target is less than 30$^\circ$, the points are colored. In (d)-(f) and (m)-(o) the targets are from the references\cite{Fang,pep-database}, and the same color represents the same target. In (g)-(i) and (p)-(r) the targets are all stable conformers we found with energy less than 0.5 eV above the global minimum. In (g)-(i) and (p)-(r) the color represents the energy of the nearest conformer: the darker color, the lower energy. 
  \label{picture:cysteine/B/2}}
\end{figure*}

First we analyze the VAE training process and acquired samples. The training loss, the latent-space scale, and the average energy of samples were all within reasonable values during the training, proving that the training went well for cysteine(see Figure S5 and SI). Next we analyze the latent space of cysteine. The trained VAE has two components: the encoder and the decoder. The latent-space data  generated by encoders with different $\beta$ are shown in Figure~\ref{picture:cysteine/B/2} (a)-(c).
The latent-space data is distributed uniformly as a circle in the latent spaces. 
For $\beta = 0$, low- and high-energy data are mixed. For $\beta < 0$, low- and high-energy regions start to form that become more pronounced for more negative $\beta$s.

To understand the correspondence between the eleven target cysteine conformers \cite{Fang} and the latent space, we discretized latent space on a $400\times 400$ grid, and mapped all points back to real space with the decoder.  We assigned each corresponding structure to one of the eleven conformers, if the MAE is smaller than $30^\circ$. If it is larger, the structure remains unassigned. The result is a map of islands in latent space, shown in panels (d)-(f). The total area of islands are 6.1\%, 3.7\% and 7.3\% of the latent space for $\beta=0, -1$ and $-3$. However, the mapping is not always unique, and multiple islands may map to the same target, such as Ia in Figure~\ref{picture:cysteine/B/2} (d). This suggests that structures that are similar in real space are not necessarily close in latent space.

The eleven targets are distributed within [0, 0.25 eV] from the global minimum. We repeatede the same procedure described in the last paragraph, but now use all the conformers we identified in the energy window [0, 0.5 eV] from the global minimum as references. We colored the latent space by the energies of these reference conformers and call the colored area low-energy areas. For  $\beta=0, -1$, and $-3$, the low-energy areas cover 36.2\%, 26.6\%, and 36.8\% of the latent space in Figure~\ref{picture:cysteine/B/2}(g)-(i).

\begin{figure*}[htpb]
  \centering
  \includegraphics[width=1.0\textwidth]{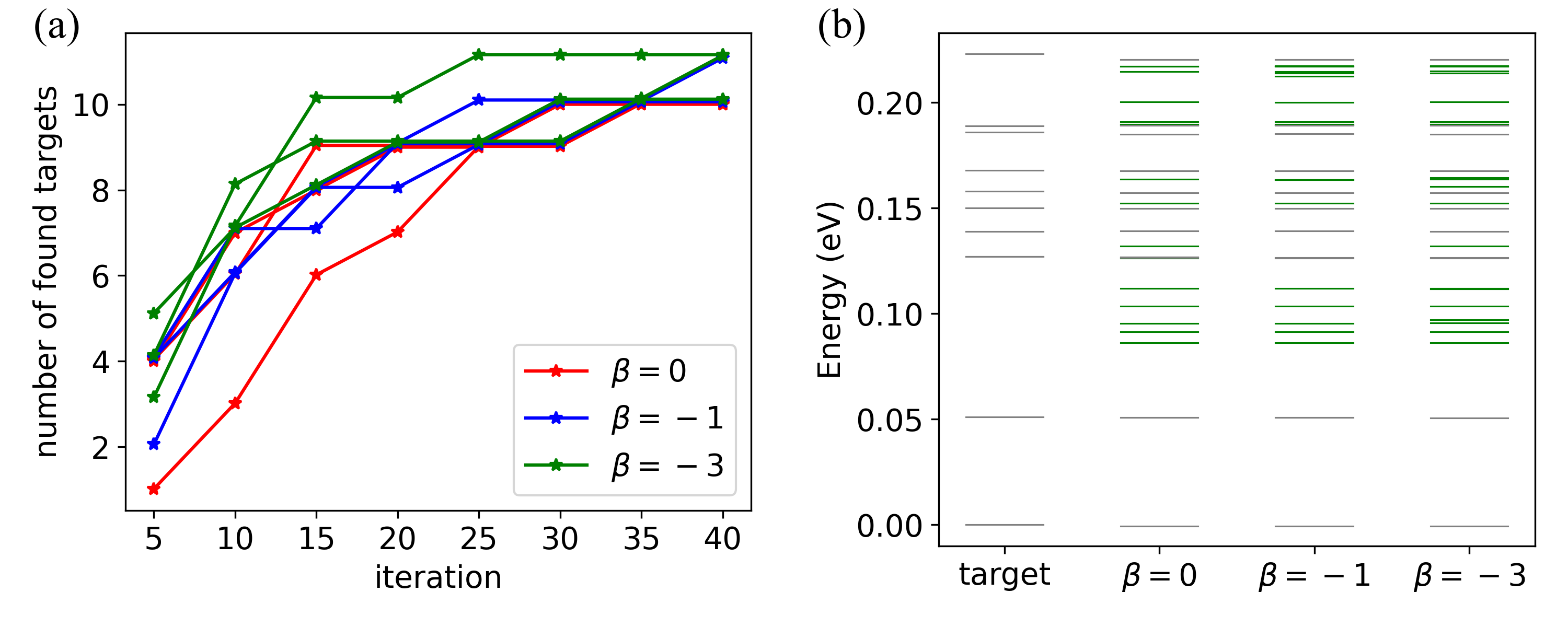}
  \caption{The results of cysteine. (a) The accumulative result for every single run. Three parallel runs are performed for each $\beta$. (b) Stable conformers found in this work.  Conformers are represented by lines and sorted by energy. The gray lines represent the found structures that are the targets from the reference \cite{Fang}. The green line represents the new conformers that are not the 11 targets. }
  \label{picture:cysteine/C/3}
\end{figure*}

Finally, we analyze and evaluate the performance of LOLS for cysteine conformer search. 
Figure~\ref{picture:cysteine/C/3}(a) shows the numbers of targets found in the nine parallel runs. The same color is used for results with the same  $\beta$ value. The $y$ value gives the accumulative number of correctly identified targets before that iteration. The best outcome is in one run with $\beta=-3$ (top green curve), while the worst result has $\beta=0$ (bottom red curve). The other seven runs perform accordingly. $\beta=-3$ runs are among the best, $\beta=-1$ average and $\beta=0$ the worst. We therefore recommend a $\beta$ value smaller than zero. The results using the same $\beta$ for three parallel runs are merged into one and shown in Figure~\ref{picture:cysteine/C/3}(b). The figure shows that all the eleven targeted conformers (along with some new ones) were found regardless of $\beta$. This is also shown again in Table~\ref{table:result}.

\subsection{GFA}

Next we applied LOLS to GFA.  The training loss, the latent-space scale and the average energy of the samples in Figure S6 indicate that the training went well for GFA. In addition, we observed that more low-energy data is generated for non-zero $\beta$s. More discussions can be found in the SI. 
We plot the latent spaces of GFA for $\beta = 0$, $-1$ and $-3$ using the same mapping methods as for cysteine. Panels (j)-(l) in Figure~\ref{picture:cysteine/B/2} show the latent-space data generated by the encoder. For $\beta=-3$, the latent-space data is more compact and contains more low-energy data than for  $\beta=0$ and $\beta=-1$. Figures~\ref{picture:cysteine/B/2}(m)-(o) show the correspondence of the latent-space data to the target GFA conformers\cite{pep-database} in real space generated by the decoder.   Unlike for cysteine, the colored latent space of GFA is quite empty. The total area of colored islands are 0.06\%, 0.28\% and 0.28\% for $\beta=0, -1$ and $-3$. Figures~\ref{picture:cysteine/B/2}(p)-(r) are colored in the same way as  Figures~\ref{picture:cysteine/B/2}(g)-(i). The low-energy areas ([0, 0,5 eV]) cover 1.0\%, 7.6\% and 11.7\% of the latent space of GFA, for $\beta = 0$, $-1$, and $-3$. The coverages are much smaller than in cysteine. We believe that this due to the higher dimensionality of GFA (9 compared to 6). Higher-dimensional systems usually have more complex PESs, and less area can be associated with low-energy conformers, which may explain the emptiness of latent space. 

\begin{figure}[htbp]
  \centering

  \includegraphics[width=0.45\textwidth]{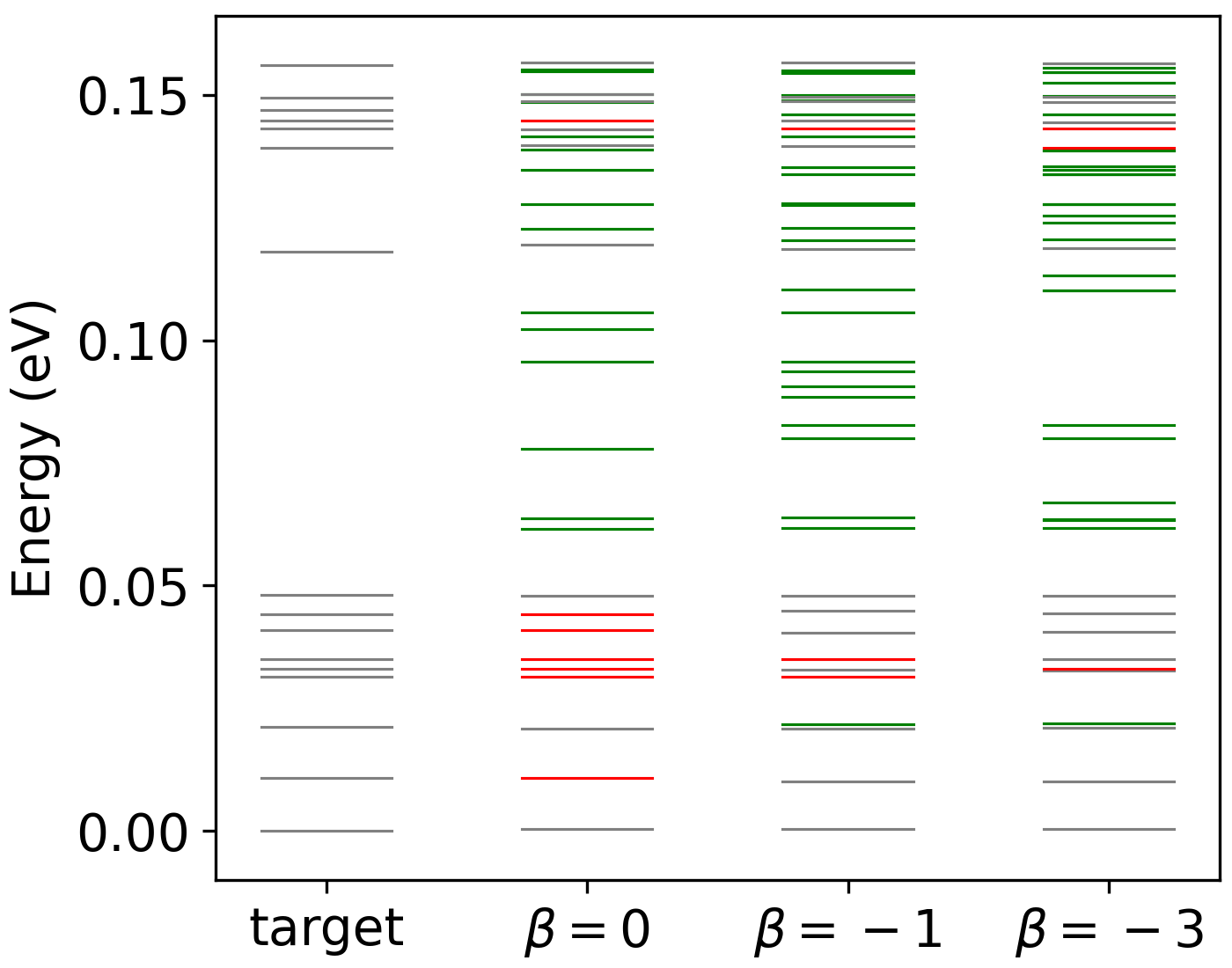}
  \caption{Stable conformers of GFA found in this work. The gray lines mean  the targets from the  reference\cite{pep-database} are found, while the red lines mean we missed the targets. The  green  line  represents  the new conformers that we found but missed by the reference. }
  \label{picture:gfa9d/C/3}
\end{figure}

For every $\beta$, the accumulative results of three parallel runs were merged into one and shown in Figure~\ref{picture:gfa9d/C/3}. We used the sixteen GFA structures reported in Ref \citenum{pep-database} as our targets. We found nine, thirteen and thirteen out of the sixteen targets for $\beta = 0, -1$ and $-3$.  Among the three values of $\beta$, $\beta=-3$ performs best for GFA, $\beta=-1$ has similar performance as $\beta=-3$, but $\beta=0$ missed six out of nine lowest energy targets. As mentioned in Ref \citenum{pep-database}, these targets can be divided into six structural types according to the different hydrogen bonds. All six types are found with $\beta$ of 0, $-1$ and $-3$.

The differences between our results and the reference results\cite{pep-database} are mainly due to the flexibility of the end groups of GFA. The \ce{-CH2NH2} branch and the \ce{-C6H6} branch (benzene ring) of GFA have several stable configurations which have energy differences within 10 meV. The two groups are at the end of the peptide, thus having little effect on the overall structures of GFA, however resulting in the different conformers. For example,  GFA 06, GFA 11, and GFA 08 have very similar structures (See Figure S7). The only difference between GFA 06 and GFA 11 is the configuration of the benzene ring, which causes an 1.7 meV energy difference. And the only difference between GFA 11 and GFA 08 is the configuration of the \ce{-CH2NH2} branch, which causes a difference of 1.9 meV. We found GFA 11 but missed GFA 06 and 08 in the result with $\beta=-1$, and we missed GFA 11 but found GFA 06 and 08 with $\beta=-3$. 
Importantly, the global minimum (GFA 15) is always found by our method even with different $\beta$s. The reference did not find any conformers in the energy range from 0.05 to 0.12 eV above the global minimum. However, we found six, ten, and eight new structures in this energy region using $\beta$ = 0, $-1$, and $-3$. Overall, we have achieved comparable accuracy as the reference.

\subsection{WG, WGG and GGF}
We also tested WG, WGG and GGF, whose search dimensions are seven, nine and nine, respectively, and compared them with Ref. \citenum{pep-database}. For each $\beta$ of 0, $-1$, and $-3$, three parallel runs were carried out for WG and GGF, with maximum iteration count $M=120$. Unfortunately, we did not find the global minimum of WGG at 120 iterations for any value of $\beta$, so we ran an additional 20 iterations  for  WGG. The accumulative results are shown in Figure S8. The results of all the five molecules are also summarized in Table~\ref{table:result}. 

For WG,  using $\beta=0$ or $-3$, we found all the thirteen targets, but $\beta=-1$ missed the highest energy target.
For GGF, the performance for different $\beta$ were close but $\beta=-1$ found the most targets.  For WGG, $\beta=0$ missed the global minimum, which was found with $\beta=-1$ or $\beta=-3$. Combined with the results for cystine and GFA, we can state that non-zero $\beta$ is at least beneficial for larger molecules such as GFA, GGF and WGG. Except cysteine, all other molecules are peptides which are very flexible molecules. It is therefore no surprise that our structure lists are not exactly the same for different $\beta$ or as the ones in Ref \citenum{pep-database}. We have missed some targets but also found some new ones in the same energy region. Overall we achieved the same level of performance as the reference.

\renewcommand{\b}{$\bullet$}
\renewcommand{\c}{$\circ$}
\begin{table*}[htp]
\begin{tabular}{|c|c|c|c|c|c|c|c|c|l|}
\hline
Material & Dim & $\beta$ & Target & Achieved & New & Single              & Relax & Converged & Achieved Details                 \\ \hline
Cysteine & 5   & 0       & 11     & 11       & 16  & 6000 & 919   & 919       & \b\b\b\b\b\b\b\b\b\b\b           \\ \hline
Cysteine & 5   & -1      & 11     & 11       & 18  & 6000 & 922   & 922       & \b\b\b\b\b\b\b\b\b\b\b           \\ \hline
Cysteine & 5   & -3      & 11     & 11       & 20  & 6000 & 944   & 944       & \b\b\b\b\b\b\b\b\b\b\b           \\ \hline
WG       & 7   & 0       & 13     & 13       & 46  & 18000 & 2314   & 2172       & \b\b\b\b\b\b\b\b\b\b\b\b\b       \\ \hline
WG       & 7   & -1      & 13     & 12       & 47  & 18000 & 2292   & 2177       & \b\b\b\b\b\b\b\b\b\b\b\b\c       \\ \hline
WG       & 7   & -3      & 13     & 13       & 45  & 18000 & 2214   & 2087       & \b\b\b\b\b\b\b\b\b\b\b\b\b       \\ \hline
GFA      & 9   & 0       & 16     & 9        & 15  & 18000 & 1443  & 1227      & \b\c\b\c\c\c\c\c\b\b\b\b\c\b\b\b \\ \hline
GFA      & 9   & -1      & 16     & 13       & 23  & 18000 & 1872  & 1588      & \b\b\b\c\b\c\b\b\b\b\b\c\b\b\b\b \\ \hline
GFA      & 9   & -3      & 16     & 13       & 27  & 18000 & 1873  & 1597      & \b\b\b\b\c\b\b\b\b\b\c\c\b\b\b\b \\ \hline
GGF      & 9   & 0       & 13     & 9        & 23  & 18000 & 1870  & 1555      & \b\b\c\b\b\b\b\b\c\b\b\c\c       \\ \hline
GGF      & 9   & -1      & 13     & 10       & 22  & 18000 & 1645  & 1381      & \b\b\b\b\b\b\b\c\b\c\b\c\b       \\ \hline
GGF      & 9   & -3      & 13     & 9        & 22  & 18000 & 1553  & 1379      & \b\b\c\b\b\b\b\c\c\b\b\c\b       \\ \hline
WGG      & 9   & 0       & 13     & 7        & 13  & 21000 & 2536  & 2004      & \c\c\b\c\b\b\b\b\c\b\b\c\c       \\ \hline
WGG      & 9   & -1      & 13     & 7        & 10  & 21000 & 3073  & 2508      & \b\c\b\c\b\c\b\b\c\b\c\c\b       \\ \hline
WGG      & 9   & -3      & 13     & 9        & 9   & 21000 & 2844  & 2270      & \b\b\b\c\b\b\b\c\c\b\c\b\b       \\ \hline
\end{tabular}

\caption{Final results for all 5 molecules. Results for three parallel run are merged. ``Achieved" means the number of targets we found. ``New" means the number of stable structures we found but missed by the reference. (Only the ones with energy less than the maximum energy of targets are counted.) ``Achieved Details" enumerate the targets sorted by energy, where $\bullet$ and $\circ$ represent found and missed targets. ``Single" means the total number of single-point energy calculations during the three parallel samplings. ``Relax" shows the number of optimized structures. ``Converged" gives the number of stable structures that are converged within 200 geometry optimization steps.}
\label{table:result}
\end{table*}

\subsection{Comparison to real space search}

\begin{figure}[htbp]
    \includegraphics[width=0.45\textwidth]{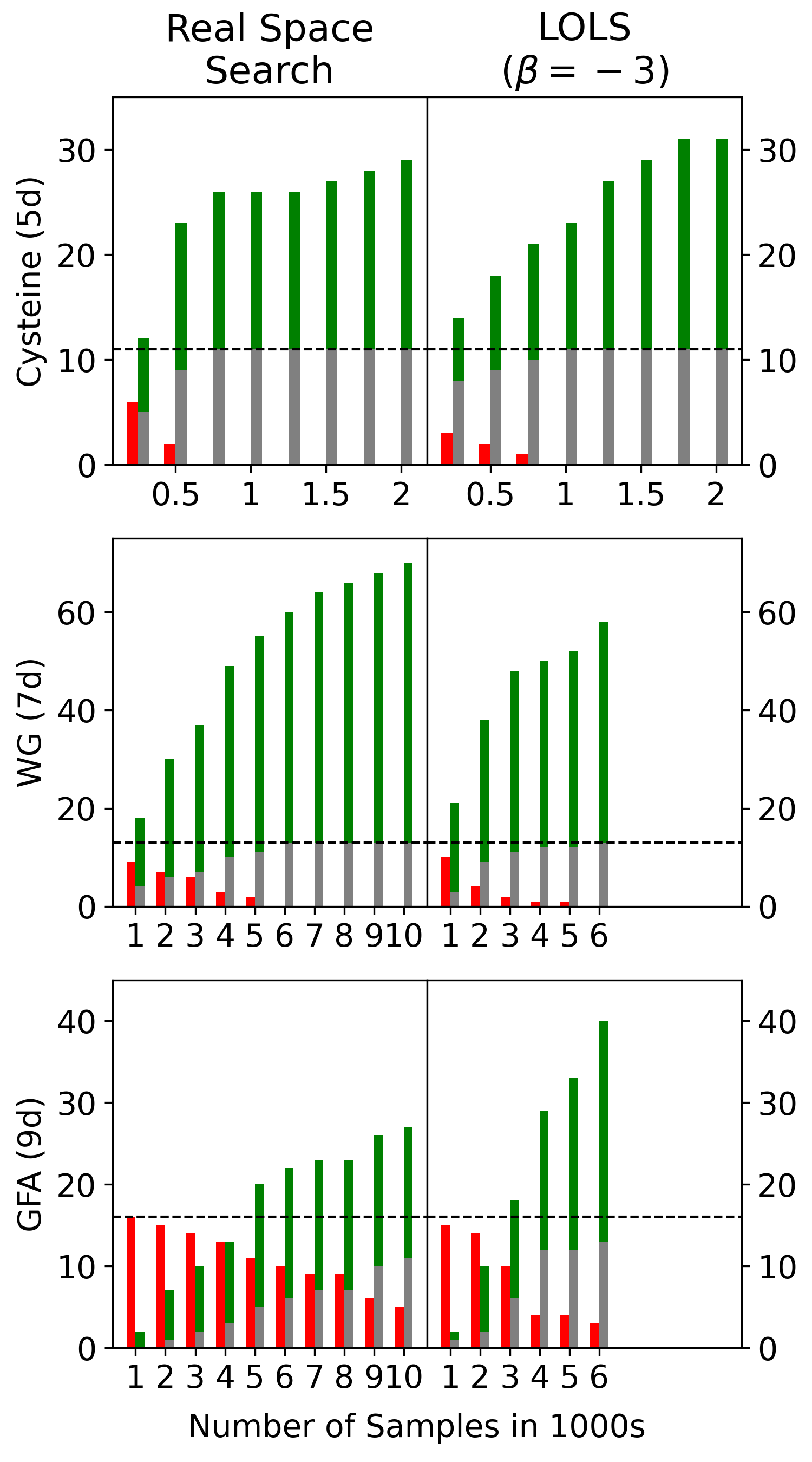}
    \caption{Results of the real space search workflow v.s. the LOLS (with $\beta=-3$) for cysteine, WG, and GFA. The x-axis is the number of samples. The height of bars represents number of conformers. Red means missing targets, gray means achieved targets and green means new conformers. The black dash lines represent the number of targets.}
    \label{picture:benchmark/all}
\end{figure}

In order to compare the sampling in low-energy latent space and in real space, we compared LOLS to a real space search workflow. In the real space search, we took samples randomly from real space and fitted a GP surrogate model every $k$ samples, gathered the local minima as the relaxation starting points, relaxed the geometries with DFT, removed duplicates and then compared them with targets (Algorithm S1). In other words, the real space search workflow replaces
the data generation loop by taking random
samples directly in real space but keeps the other steps of LOLS.
We tested the real space search workflow on cysteine(5-D), WG (7-D), and GFA(9-D). For each molecule, we carried out three parallel runs. The results of the parallel runs were merged and compared to LOLS with $\beta=-3$ in Figure~\ref{picture:benchmark/all}. The details of the observed targets are shown in Figure S9.

Figure~\ref{picture:benchmark/all} presents the number of targets found versus the number of samples used to build the energy models.  For cysteine, the real space search  found all the eleven targets with 2250 samples, while LOLS ($\beta=-3$) required 3000 samples. For WG, the real space search and LOLS both took 18000 samples to find all the thirteen targets.  LOLS's performance is similar to the real space search for cysteine and WG. However, for GFA LOLS starts to provide an advantage. The real space search found eleven out of sixteen targets using 30000 samples, while LOLS ($\beta=-3$) found twelve targets with 12000 samples and thirteen targets with 18000 samples. LOLS clearly outperforms the real space search, indicating our method is more suitable for larger molecules with more degrees of freedom.

\subsection{Discussion}

First, we discuss the properties of latent spaces in this work. Our analysis of the 2-D latent spaces generated by encoders revealed them to be neither smooth nor continuous (See Figure~\ref{picture:cysteine/B/2}(a)-(c) and (j)-(l)). High- and low-energy areas appear intermixed in the latent space, and it proved difficult to fit GP models to latent-space data and extract any information on low energy regions. Moreover, casting previously known conformers into latent space demonstrated that the same conformer structure can be mapped into different locations in latent space (see Figure~\ref{picture:cysteine/B/2}(d)-(i) and (m)-(r)). This suggests that similar structures in real space are not necessarily close in latent space. For these reasons, we did not further pursue designing acquisition functions or minima searches in latent space. Instead we use the fast, explorative and space-filling random sampling approaches to sample  latent space.

We analyzed the low-energy area ([0, 0.5 eV]) of our latent spaces. For cysteine, all workflows with the different $\beta=0,-1,-3$ achieved good results, which may be due to the similar coverage of low-energy area ($\sim$ 30\%).   However, for GFA,  only 1\% of latent space corresponds to low-energy structures for $\beta = 0$, which is likely to be the reason for missing most of the targets (See Figure 8). This percentage increases to 10\% for $\beta = -1$ and $-3$, and we achieved much better results. This analysis suggests that a non-zero $\beta$ is an advantage for LOLS. More detailed discussion of how the $\beta$ affects the datapool can be found in SI (Figure S10 and Figure S11).

Next, we discuss the efficiency of LOLS. Building a high-dimensional energy model and thoroughly exploring it requires a large amount of data. For example, if we take the grid sampling method in nine-dimensional space and divide each dimension into ten equal parts, we would need $10^9$ samples. Although our work does not aim to achieve the highest efficiency, we acquired enough data to build a reliable energy model for nine-dimensional peptides with 18000 - 21000 single-point energy calculations. We compared LOLS to a real space sampling algorithm, and conclude that LOLS found more conformers with fewer samples than the real space search algorithm for 9-D molecules. However,  for small molecules such as cysteine (5-D) and WG (7-D), our method is unlikely to outperform this real space sampling.

The LOLS algorithm is flexible and offers several  parameters that could further be modified and optimized for different systems. Also, we chose the rectangle random sampling method for simplicity. Replacing it with some more sophisticated sampling methods may further increase the efficiency. The architecture of the neural networks in the VAE, and the hyperparameters $\beta$ and $\lambda$ could also be fine-tuned for even better performance. After acquiring a fixed amount of samples, we use a GP model to build the energy model and gather the local minima for post-relaxation. Here the GP model could be replaced by any continuous model, for example, a neural network.

\section{Conclusion}

In this work, we have developed the active learning workflow LOLS for molecular conformer search. LOLS is a stochastic method that contains two machine learning models: the generative model VAE for data sampling and the GP for energy model fitting. We introduced the hyperparameter $\beta$  to steer the latent space towards low-energy molecular configurations for generating more informative data. We have applied LOLS to cysteine and the peptides WG, GFA, GGF, and WGG, and achieved a similar level of accuracy as the references. For small molecules such as cysteine, it is more efficient to sample data in real space; however, LOLS is more suitable for larger molecules such as peptides. LOLS is still at an early stage of development: further optimization of the generative model and energy model may increase the efficiency and facilitate applications to other systems beyond molecules. 

We have also gained insight into the nature and properties of latent space both quantitatively and qualitatively. Quantitatively, we found that the distribution of latent-space data can be controlled by the hyperparameter $\lambda$ that is used to balance the reconstruction loss and regulation term in the loss function of the VAE. By tuning $\lambda$, a more uniform latent space can be formed, which is beneficial for sampling. In addition, we found that the latent-space scale ($L$) is a good parameter to measure the size of latent space. Qualitatively, we found for cysteine and GFA that latent space is neither smooth nor continuous in the low-energy regions. Moreover, the structures are close in real space might not be close in latent space.
Therefore we recommend exploratory and space-filling sampling approaches for latent space sampling.

\subsection{References}

\begin{acknowledgement}
X.~G, L.~F. and X.~C. acknowledge the financial support from the Academy of Finland (project numbers 308647, 314298, 335571). X.~G, Y.~X. and W.~D. acknowledge the financial support from the Basic Science Center Project of NSFC (Grant No. 51788104), the National Science Fund for Distinguished Young Scholars (Grant No. 12025405), the National Natural Science Foundation of China (Grant No. 11874035), and the Beijing Advanced Innovation Center for Future Chip (ICFC), M.~T. and P.~R. have received funding from the Academy of Finland via the Artificial Intelligence for Microscopic Structure Search (AIMSS) project No.~316601 and the Flagship programme: Finnish Center for Artificial Intelligence FCAI. 
X.~G. and X.~C Generous computational resources were provided by CSC -- IT Center for Science, Finland, and the Aalto Science-IT project. L.~F also acknowledges financial support from the Chinese Scholarship Council (grant no. [2017]3109).

\end{acknowledgement}

\begin{suppinfo}

(Figure S1) The distributions of dihedral angles and scaled energy of the CYS800 data set; (Figure S2) the progression of training loss with training epochs; (Figure S3) the latent-space data distributions of the CYS800 data set with different $\lambda$; (Figure S4) the targets found at different iterations; (Figure S5 and S6)  the latent-space scale and the energies of samples 
during the data generation step for cysteine and GFA; (Figure S7) three very similar GFA conformers: GFA 08, GFA 11 and GFA 06.; (Figure S8) the accumulative results for WG, GGF and WGG; (Figure S9) comparison of LOLS and real space search workflow on GFA; (Figure S10) the relationship between the reconstruction error and the scaled energy in the last iteration of LOLS on cysteine; (Figure S11) the energy distribution of the data in the last LOLS iteration for cysteine, WG and GFA.

\end{suppinfo}

\bibliography{achemso-demo}

\end{document}


\section{Analysis of CYS800 data set (CYS800)}
\begin{figure}[htp]
    \centering
    \includegraphics[width=0.8\textwidth]{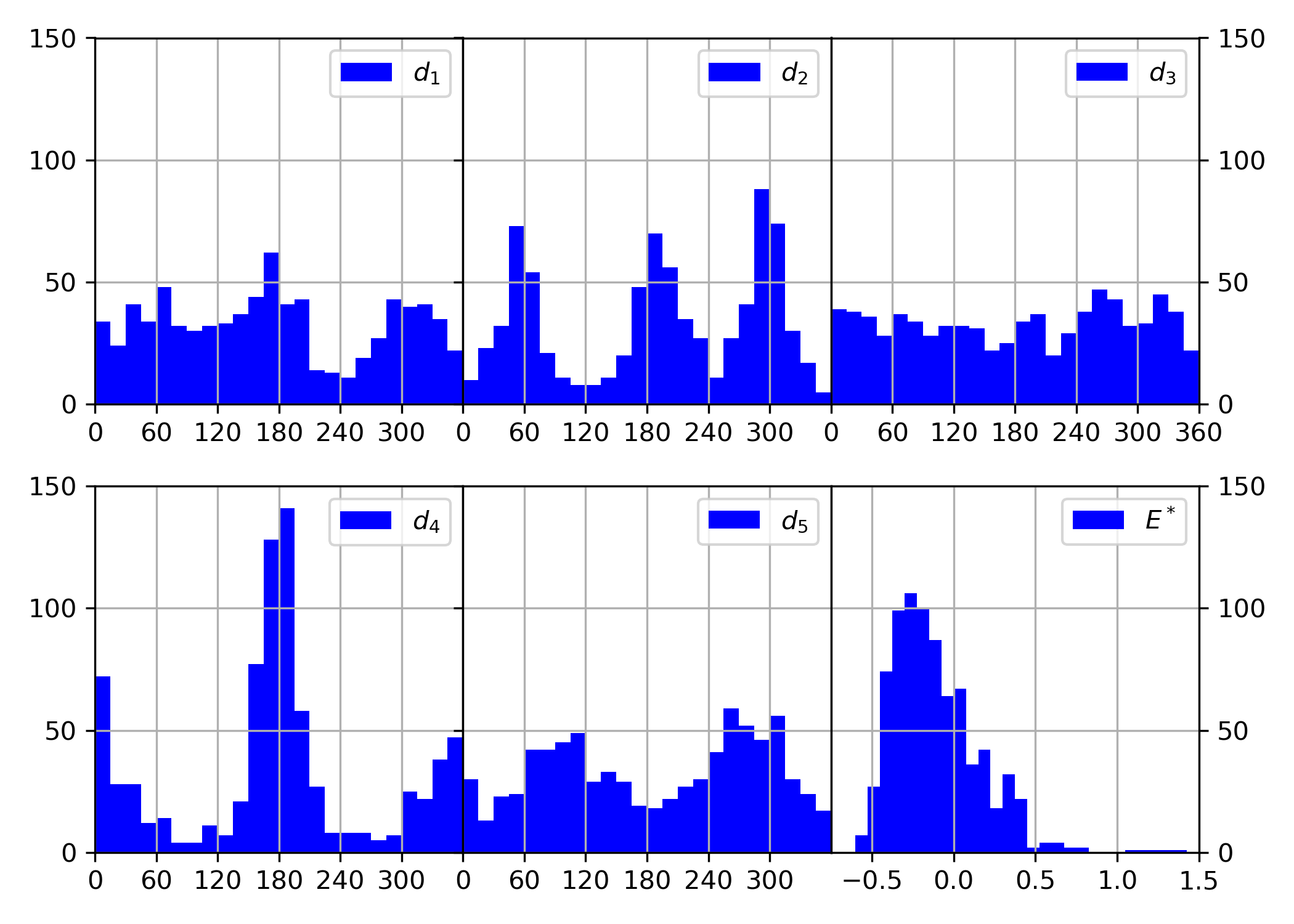}
    \caption{The dihedral angles and energy distribution of test set.}
    \label{figure:testset}
\end{figure}

Figure \ref{figure:testset} shows the distributions of dihedral angles and scaled energy of the CYS800 data set. The size of the data set is 800. The histogram shows that $d_1$ and $d_3$ are uniformly distributed in [0$^{\circ}$, 360$^{\circ}$], but the distributions of $d_2$, $d_4$, and $d_5$ have three, two and two peaks. Only 5 data in the test set has the scaled energy above 1.0. And most data has energy around the peak at $-0.25$.

\section{The progression of training loss}
\begin{figure}[htp]
    \centering
    \includegraphics[width=0.6\textwidth]{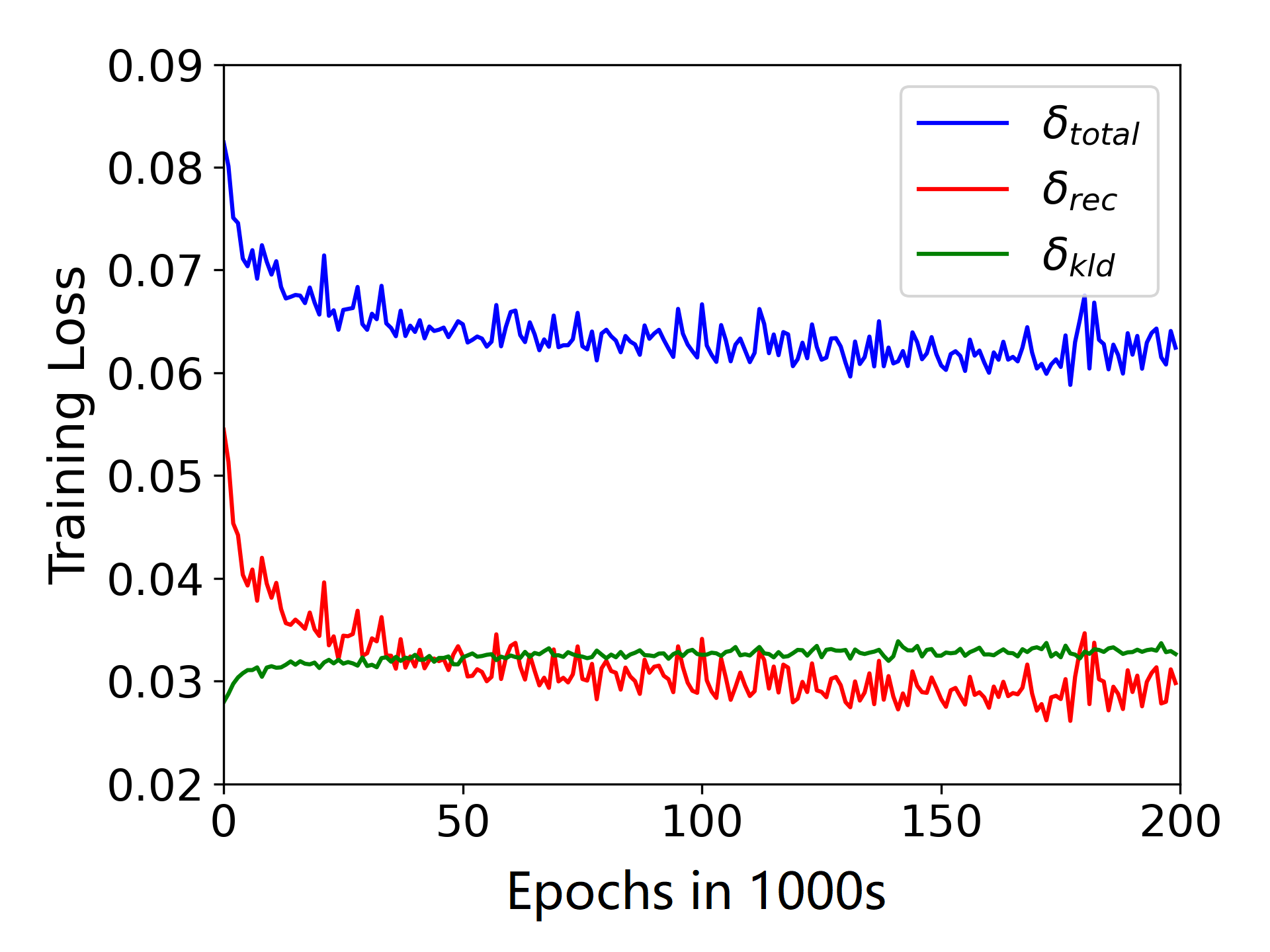}
    \caption{The progression of training loss with training  epochs.}
    \label{figure:vaeloss}
\end{figure}

Figure \ref{figure:vaeloss} shows the training loss of the CYS800 data set  changes with training epochs.  The VAE models are trained by using $\lambda=0.01, \beta=0$ and $layersize=128$. The blue, red and green lines represent the total loss, the reconstruction loss, and the KL divergence. The KL divergence keeps stable after 10000 epochs. The reconstruction loss decreases to 0.03 at the first 50000 epochs and stabilizes after 100000 epochs. We select 100000 epochs to ensure the convergence of the network training.

\section{The  latent-space data distributions with different $\lambda$ }
\begin{figure}
    \includegraphics[width=0.9\textwidth,trim=20 0 10 10,clip]{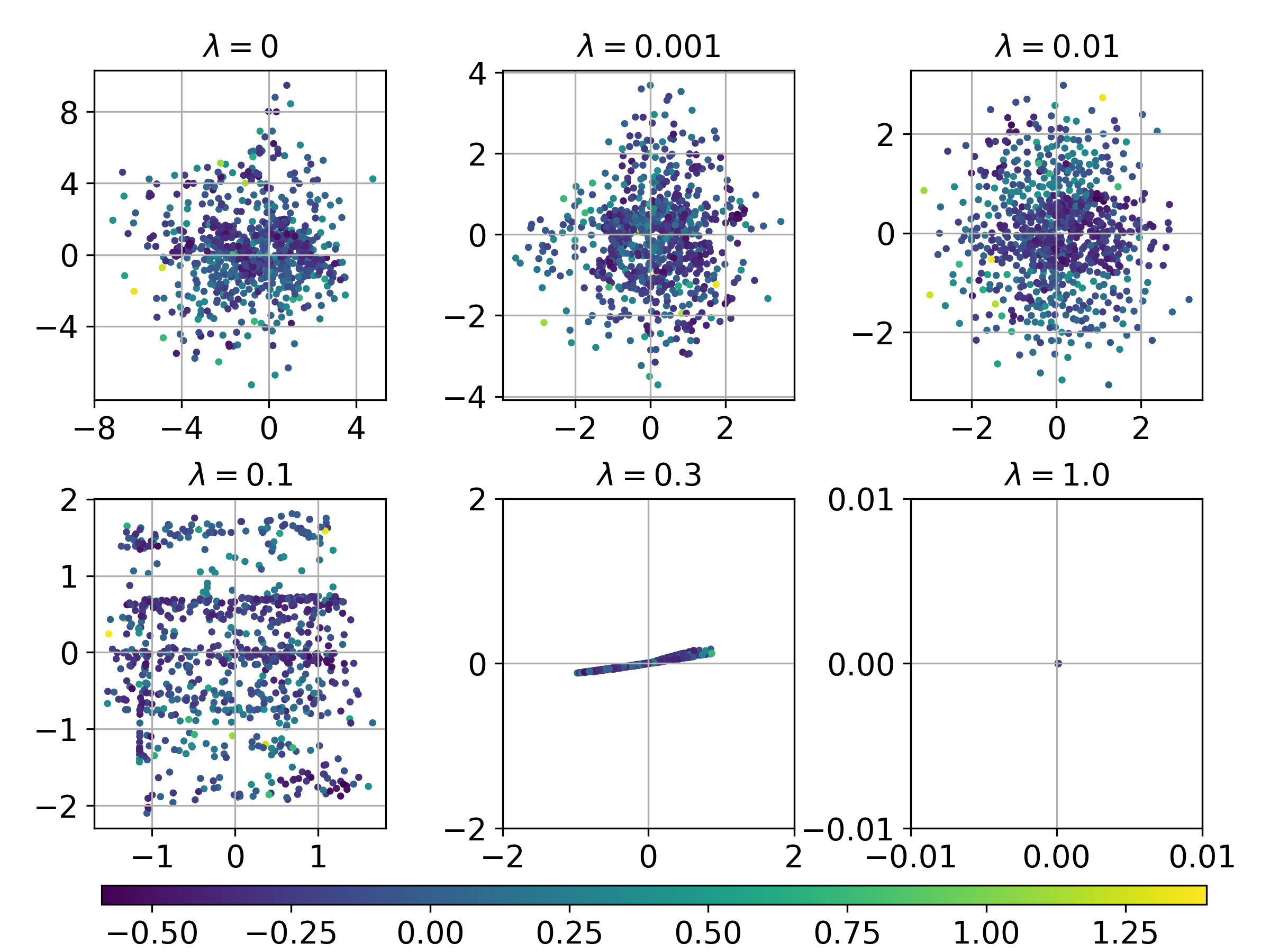}

    \caption{The latent-space data distributions of the CYS800 data set with different $\lambda$.}
    \label{figure:latentspacelambda}
\end{figure}

Figure \ref{figure:latentspacelambda} shows the latent-space data distributions of the CYS800 data set with different $\lambda$ values. We plot the latent-space data ($\mu_{i1}$, $\mu_{i2}$) and the color represents the scaled energy of the corresponding molecular structure. For $\lambda = 0$, the VAE decays to a normal auto-encoder and the data distribution is non-uniform: data points in the center of latent space are closed to each other, while other data points are sparsely distribute in the rest of latent space. When $\lambda$ increases a small amount to 0.001, the scale of latent space drops (see axis labels in Figure \ref{figure:latentspacelambda}). When $\lambda=0.01$, the data distribute even more uniformly inside a circle. When $\lambda$ increases to 0.1 the latent space restructures into parallel lines. Further increasing $\lambda$ compresses latent space and it finally collapses into a single point for $\lambda \sim 1$.

\section{Statistic analysis of cysteine results}
\begin{figure}[htp]
    \centering
    \includegraphics[width=0.8\textwidth]{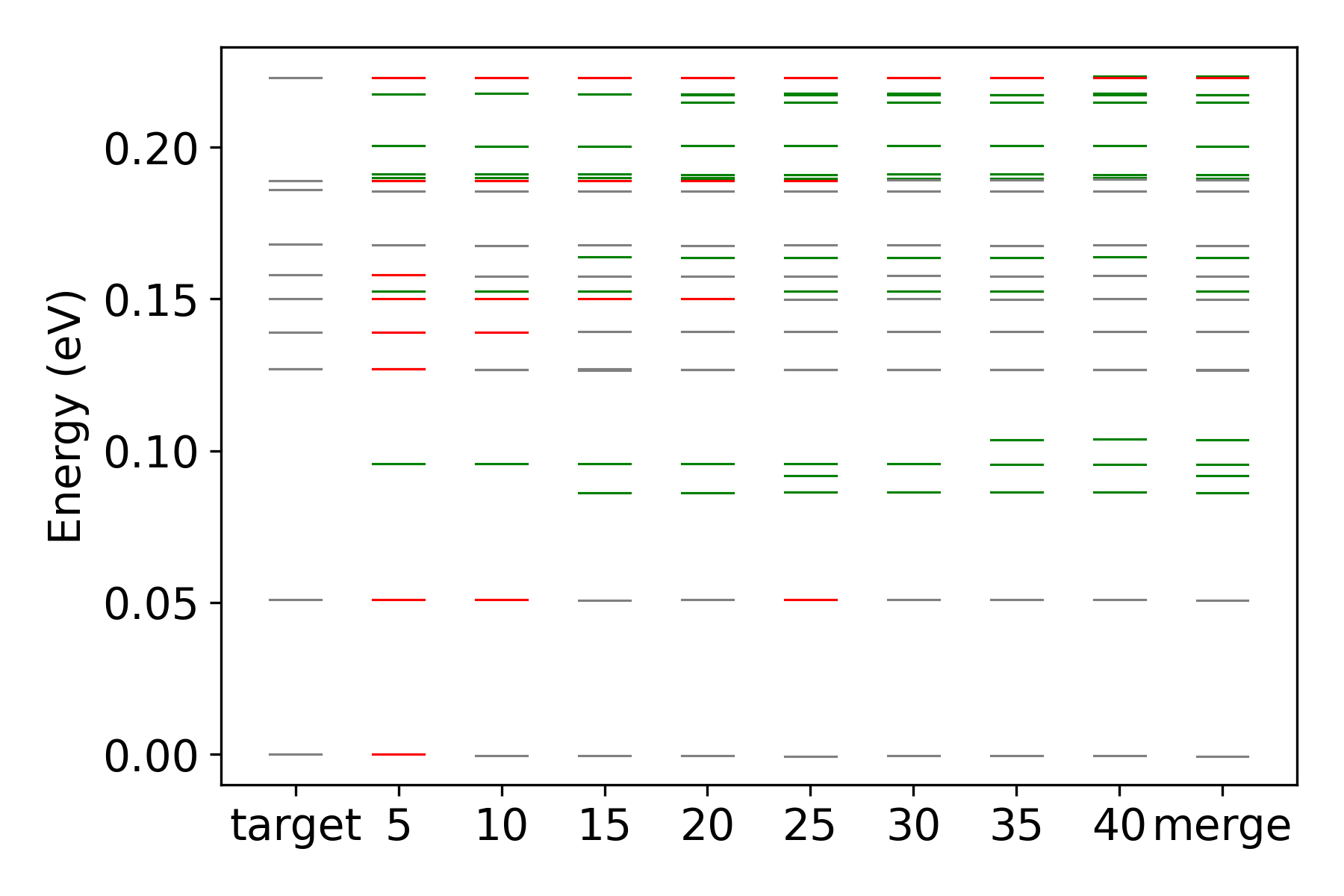}
    \caption{The targets found at different iterations.}
    \label{figure:targetmiss}
\end{figure}

We extract the local minima from the GP at different iterations and use them to initialize DFT geometry optimization. The candidate structures are further optimized with DFT structure relaxation and the results are shown in Figure \ref{figure:targetmiss}. Some targets found at certain iteration might be missed in following iterations. For example, the second target (sorted by energy) is found at iteration 15 and 20 but missed at iteration 25. This suggests us not to keep only the result of the final energy model but  the accumulative results.

\begin{figure}[htp]
    \centering
    \begin{overpic}[width=0.325\textwidth]{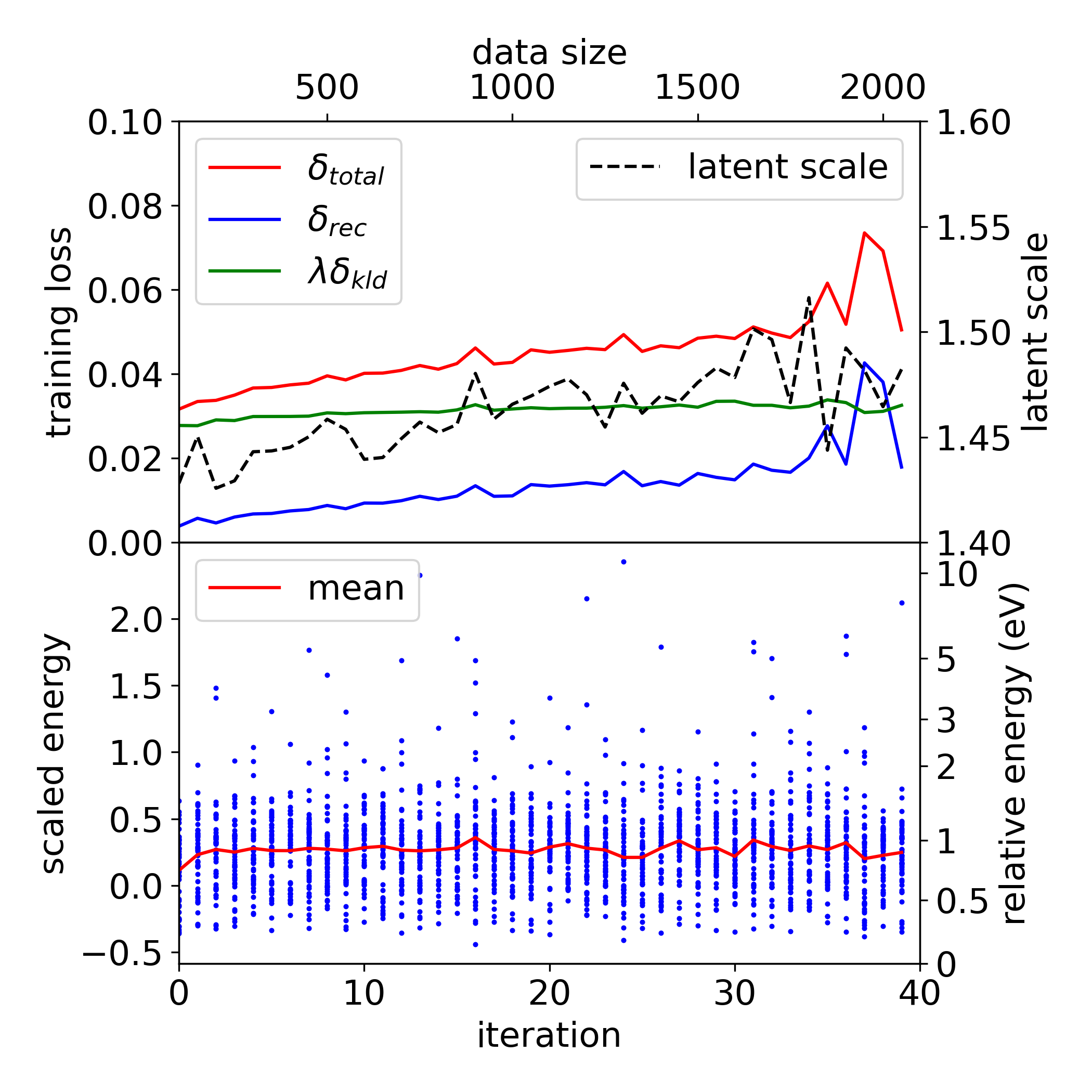}
        \put(0, 95){\textbf{(a)}$\beta = 0$}
    \end{overpic}
    \begin{overpic}[width=0.325\textwidth]{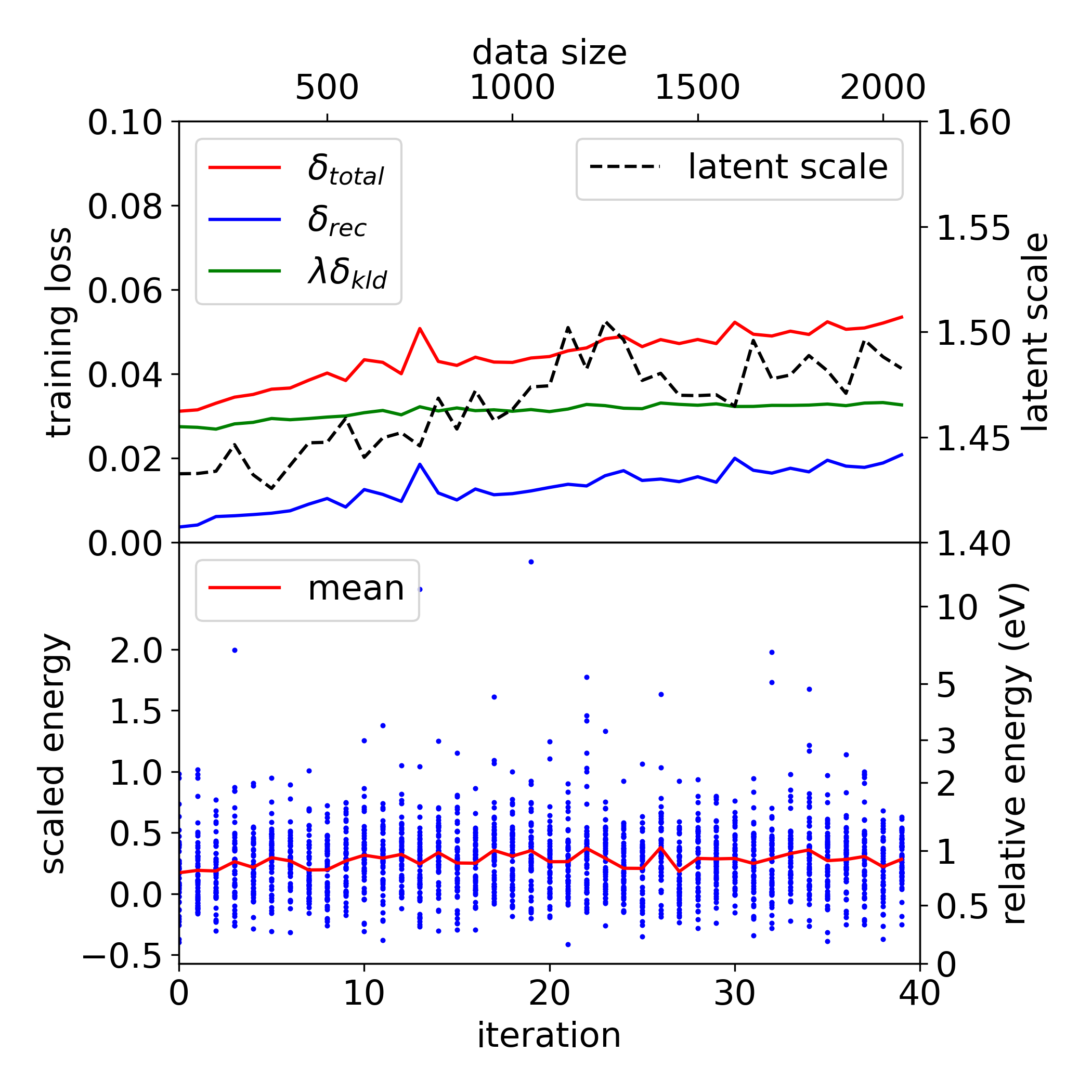}
        \put(0, 95){\textbf{(b)}$\beta = -1$}
    \end{overpic}
    \begin{overpic}[width=0.325\textwidth]{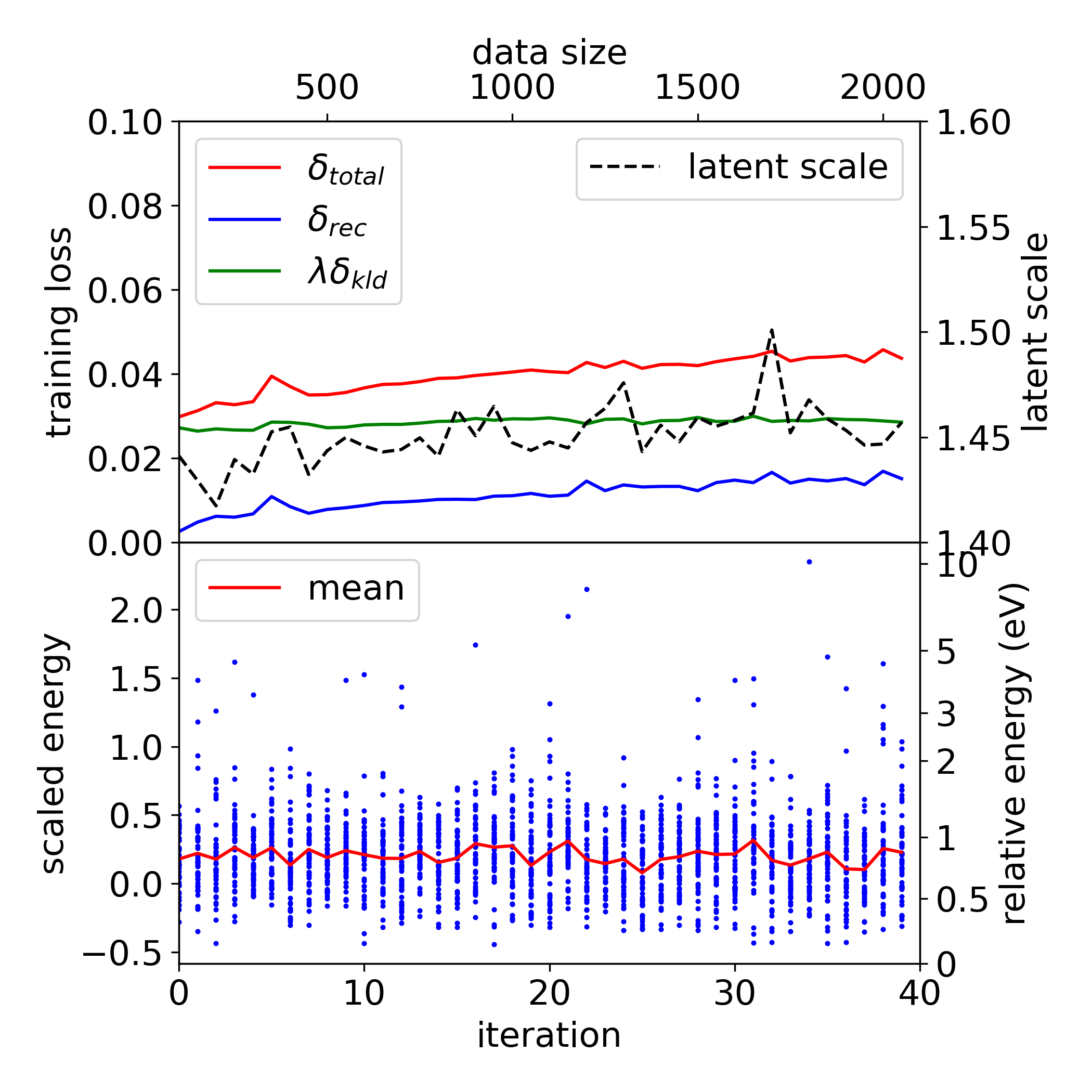}
        \put(0, 95){\textbf{(c)}$\beta = -3$}
    \end{overpic}
    
    \caption{The training loss, latent-space scale, and energies of samples  during the data generation. (a)$\beta = 0$ (b)$\beta = -1$ (c) $\beta = -3$. }
    \label{figure:cysteine3in1}
\end{figure}

Figure \ref{figure:cysteine3in1} shows how the training loss, the latent-space scale and the energies of samples changes during the data generation step for cysteine. For the same $\beta$, the KL Divergence remains stable among iterations because it is controlled by the hyperparameter $\lambda$. The fixed $\lambda$ also limits the value of latent-space scale $L$ in the range $1.40 -1.55$, which is closed to the latent-space scale $L\sim1.47$ of the test set. The reconstruction loss and the total loss increase as the iteration increases, but the increasing rate and the start value are not related to the values of $\beta$. 

\section{Statistic analysis of GFA results}

\begin{figure}[htp]
    \centering
    \begin{overpic}[width=0.325\textwidth]{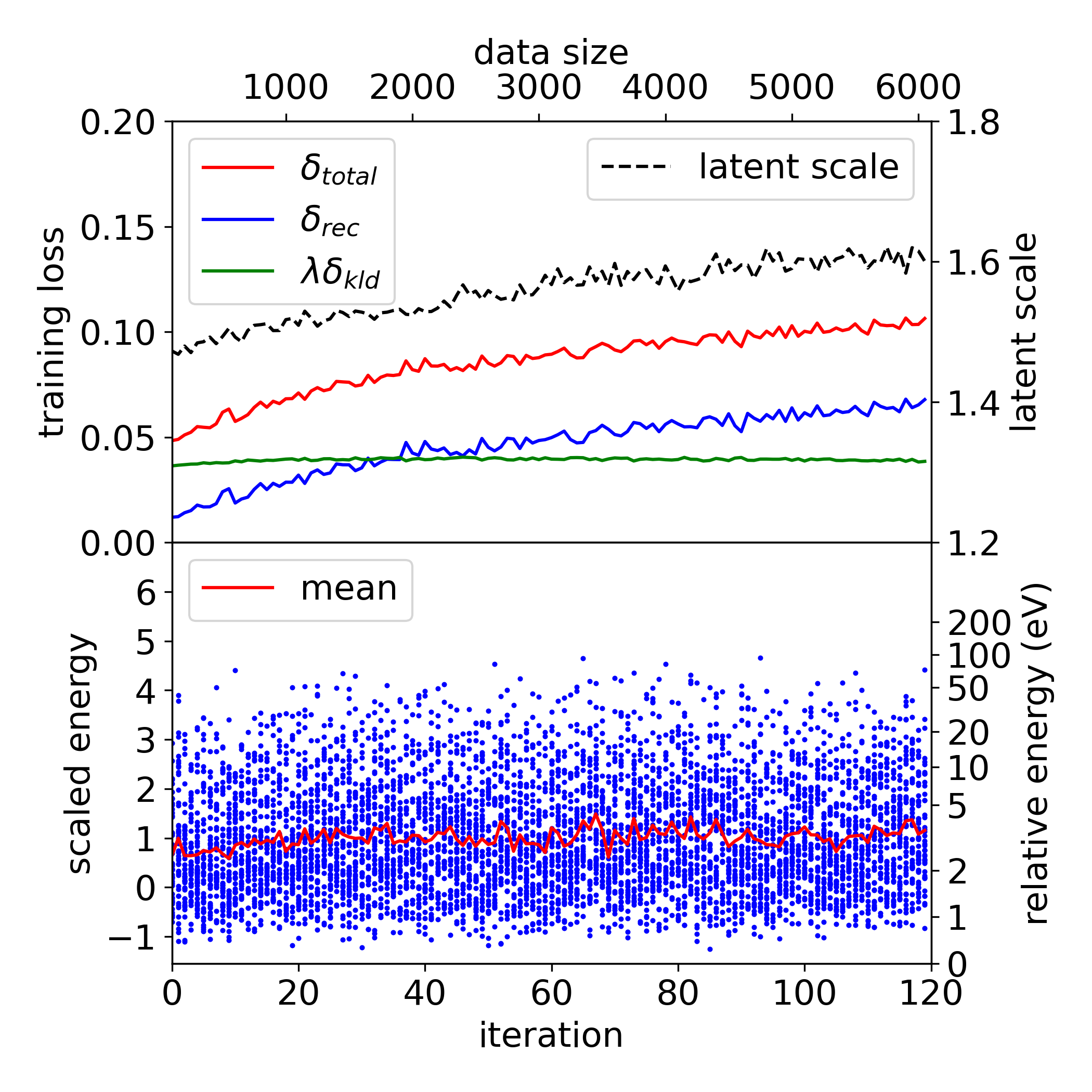}
        \put(0, 95){\textbf{(a)}$\beta = 0$}
    \end{overpic}
    \begin{overpic}[width=0.325\textwidth]{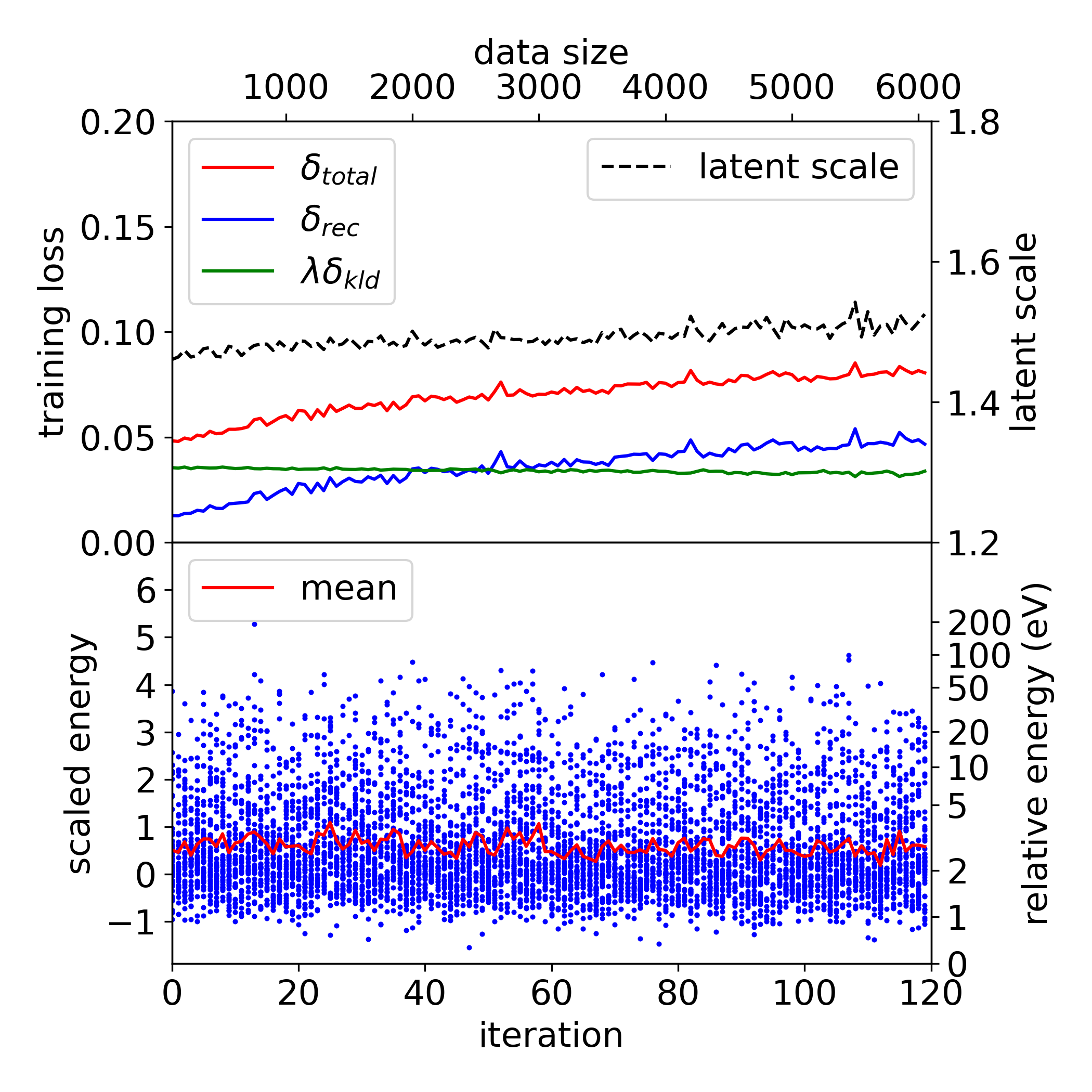}
        \put(0, 95){\textbf{(b)}$\beta = -1$}
    \end{overpic}
    \begin{overpic}[width=0.325\textwidth]{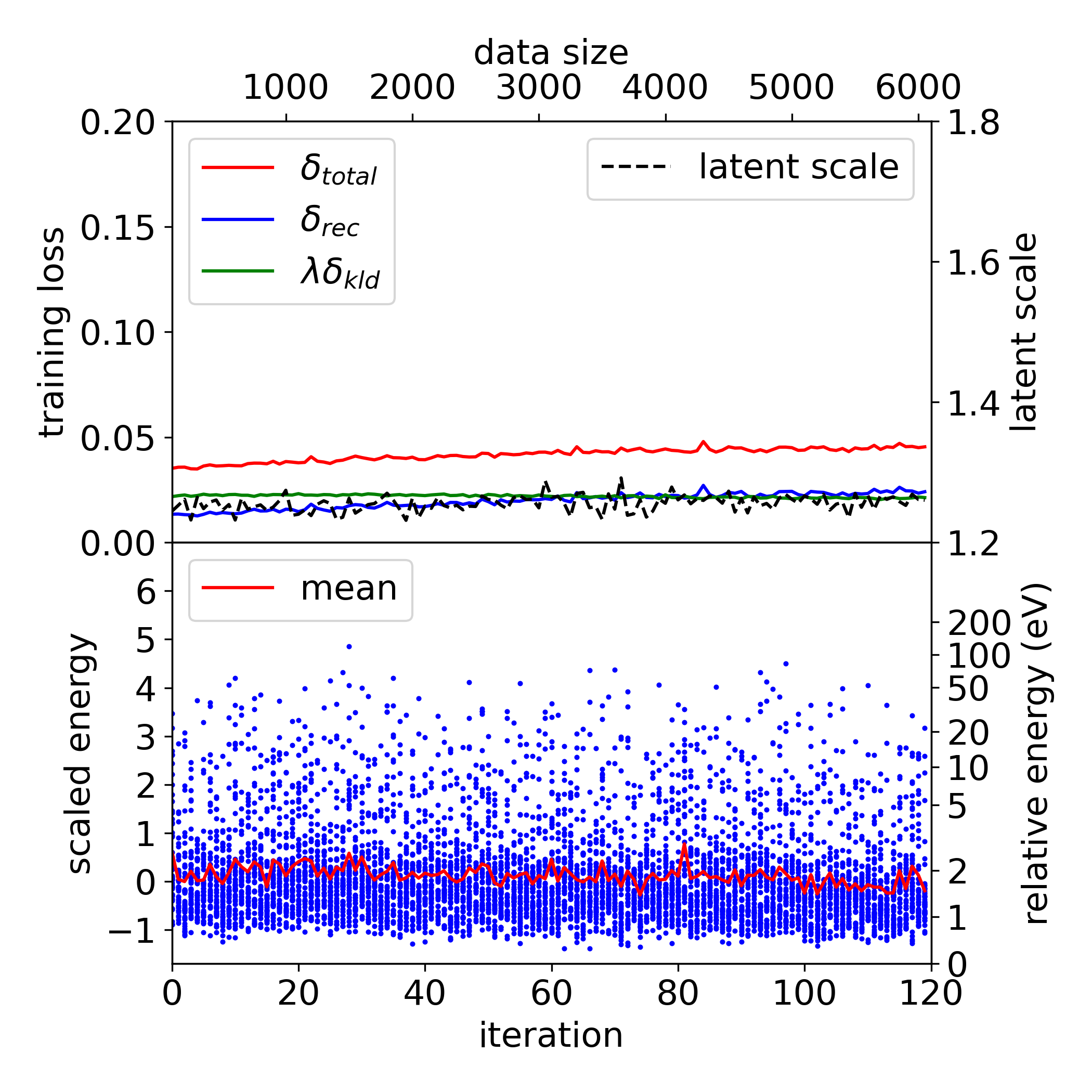}
        \put(0, 95){\textbf{(c)}$\beta = -3$}
    \end{overpic}
    
    \caption{The training loss, the latent-space scale, and the energies of samples during the data generation. (a)$\beta = 0$ (b)$\beta = -1$ (c) $\beta = -3$.}
    \label{figure:gfa3in1}
\end{figure}

Figure \ref{figure:gfa3in1} shows the reconstruction loss, the latent-space scale and the energies of samples during the data generation step for GFA . For each run, both the reconstruction loss and the latent-space scale slowly increase as the iteration steps increase, in which the workflow with $\beta=-3$ has a minimal increase rate. The KL-Divergence has a stable value among iterations, but the value for $\beta=-3$ is only half of which for $\beta=0$ or -1. The latent-space scale $L$ is about 1.25 for $\beta=-3$, which is smaller than $1.47-1.61$ for $\beta=0$ and $1.45-1.52$ for $\beta=-1$.  The average energy of samples has $\sim10\%$ fluctuation among iterations but decreases about 1 eV when $\beta$ changes from 0 to $-3$. More low-energy samples are acquired for $\beta=-3$ than $\beta=-1$ or 0.

\section{Similar GFA conformers}
\begin{figure}[htp]
    \centering
    \includegraphics[width=0.8\textwidth]{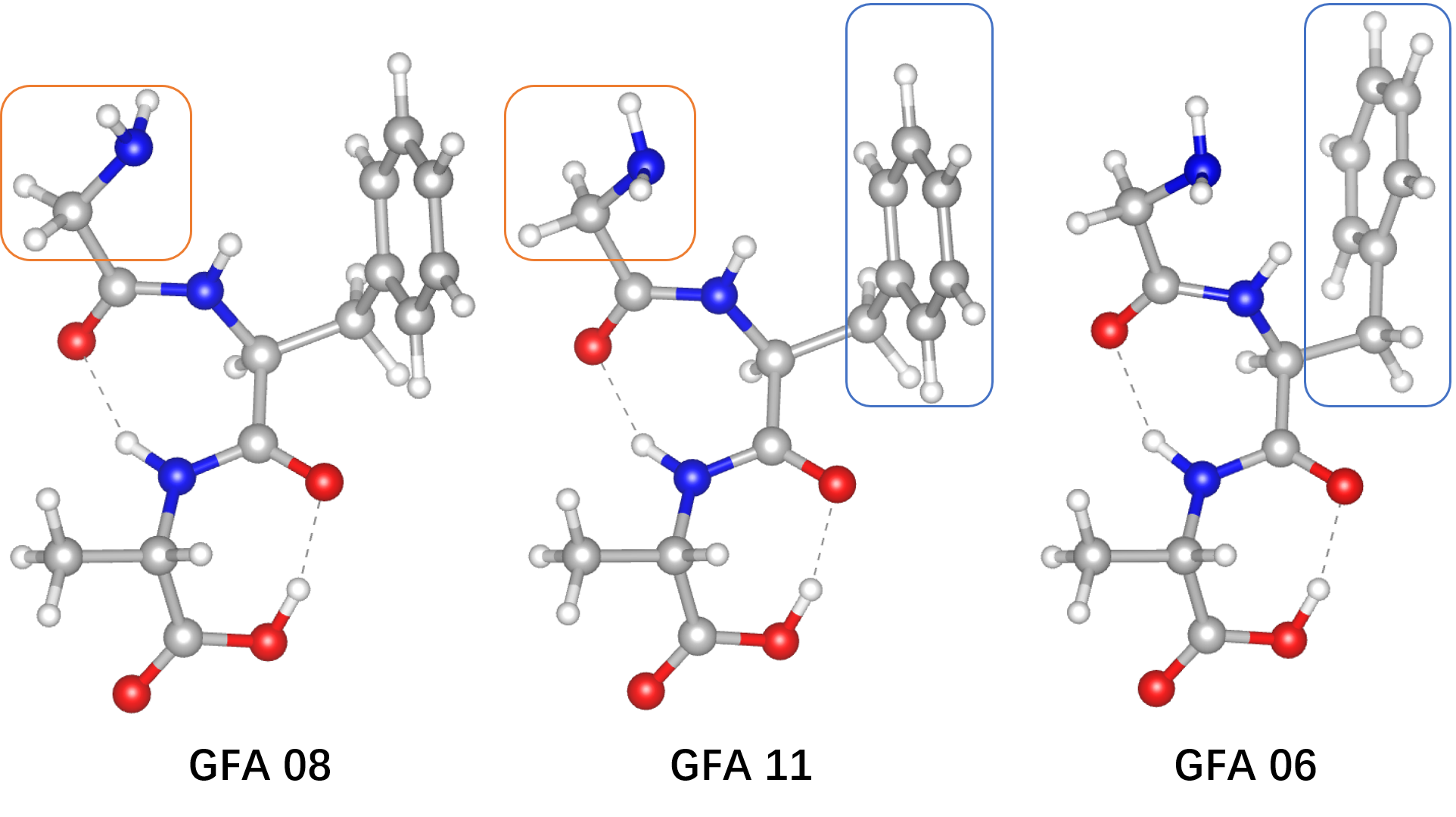}
    \caption{Three very similar GFA conformers: GFA 08, GFA 11 and GFA 06. The orange rectangles show the main difference between GFA 08 and GFA 11. The blue rectangles show the main difference between GFA 11 and GFA 06.}
    \label{figure:gfax3}
\end{figure}

Figure \ref{figure:gfax3} shows three very similar GFA conformers. The blue rectangles show the difference between GFA 06 and GFA 11 is at the \ce{-C6H6} branch (benzene ring), which causes the energy difference 1.7 meV. The orange rectangles show the difference between GFA 11 and GFA 08 is at the \ce{-CH2NH2} branch, which causes the energy difference 1.9 meV.

\section{Detailed results of WG, GGF and WGG}

\begin{figure}[htp]
    \centering
    \begin{overpic}[width=0.45\textwidth]{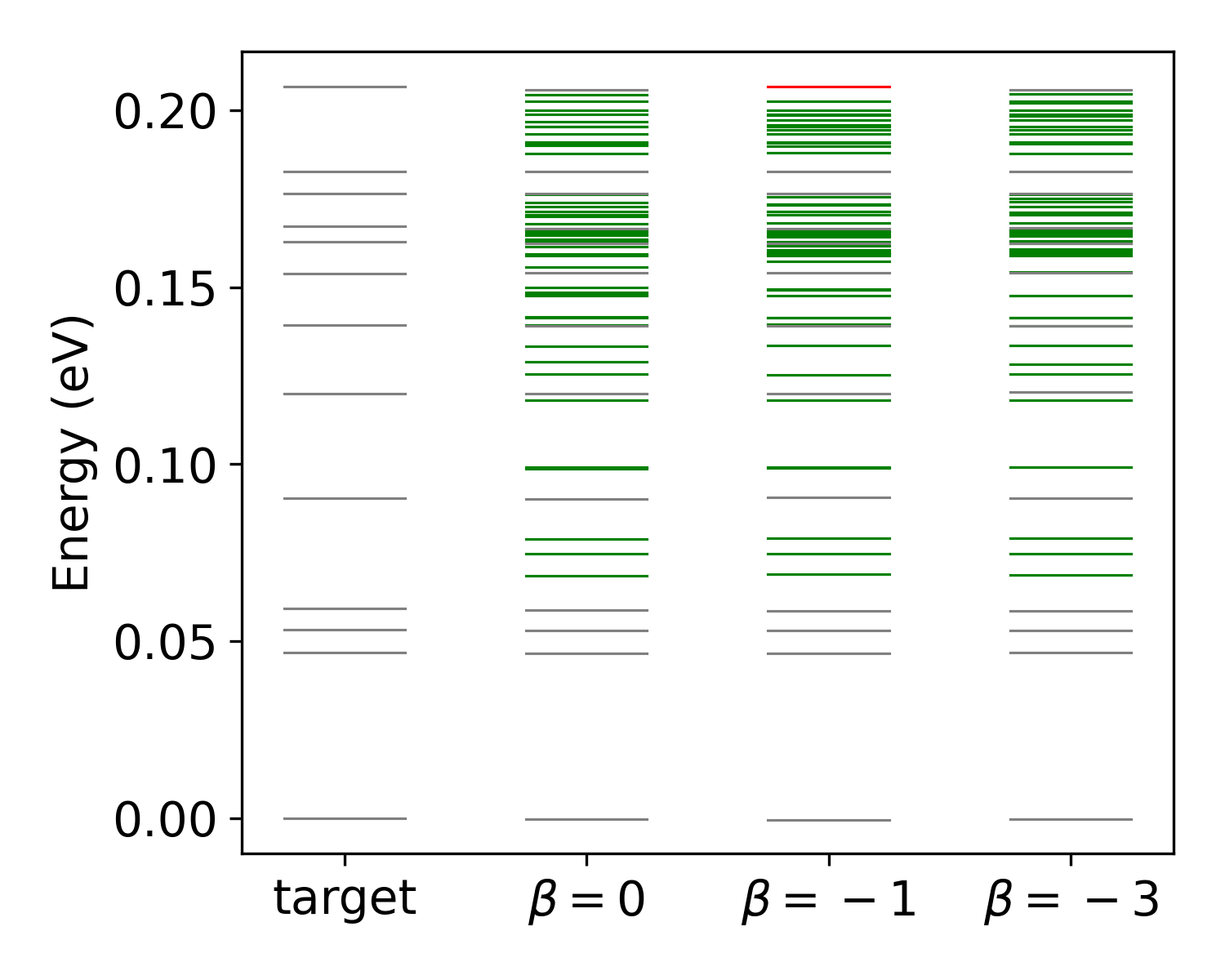}
        \put(0, 78){\textbf{(a) WG}}
    \end{overpic}
    \begin{overpic}[width=0.45\textwidth]{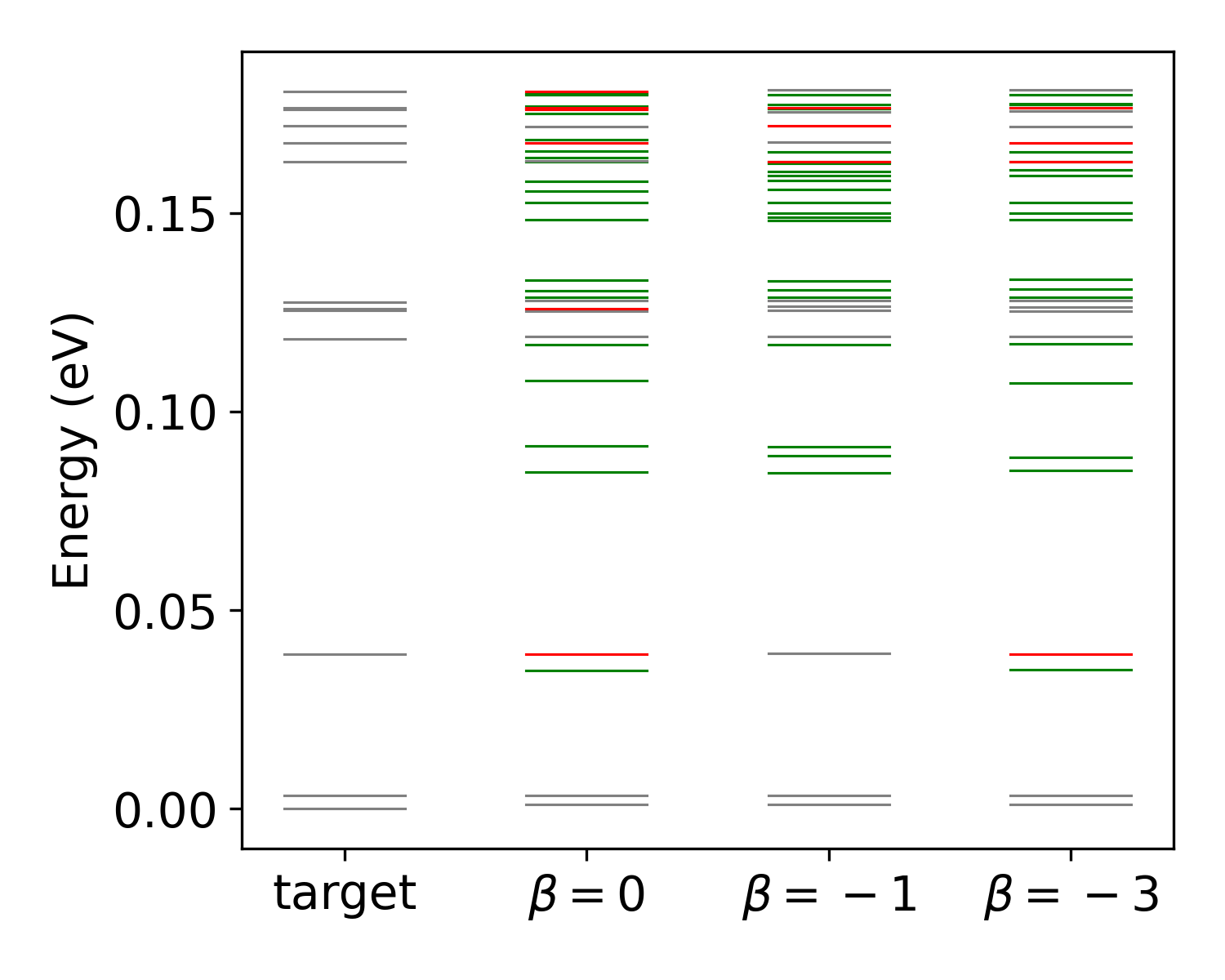}
        \put(0, 78){\textbf{(b) GGF}}
    \end{overpic}
    \begin{overpic}[width=0.45\textwidth]{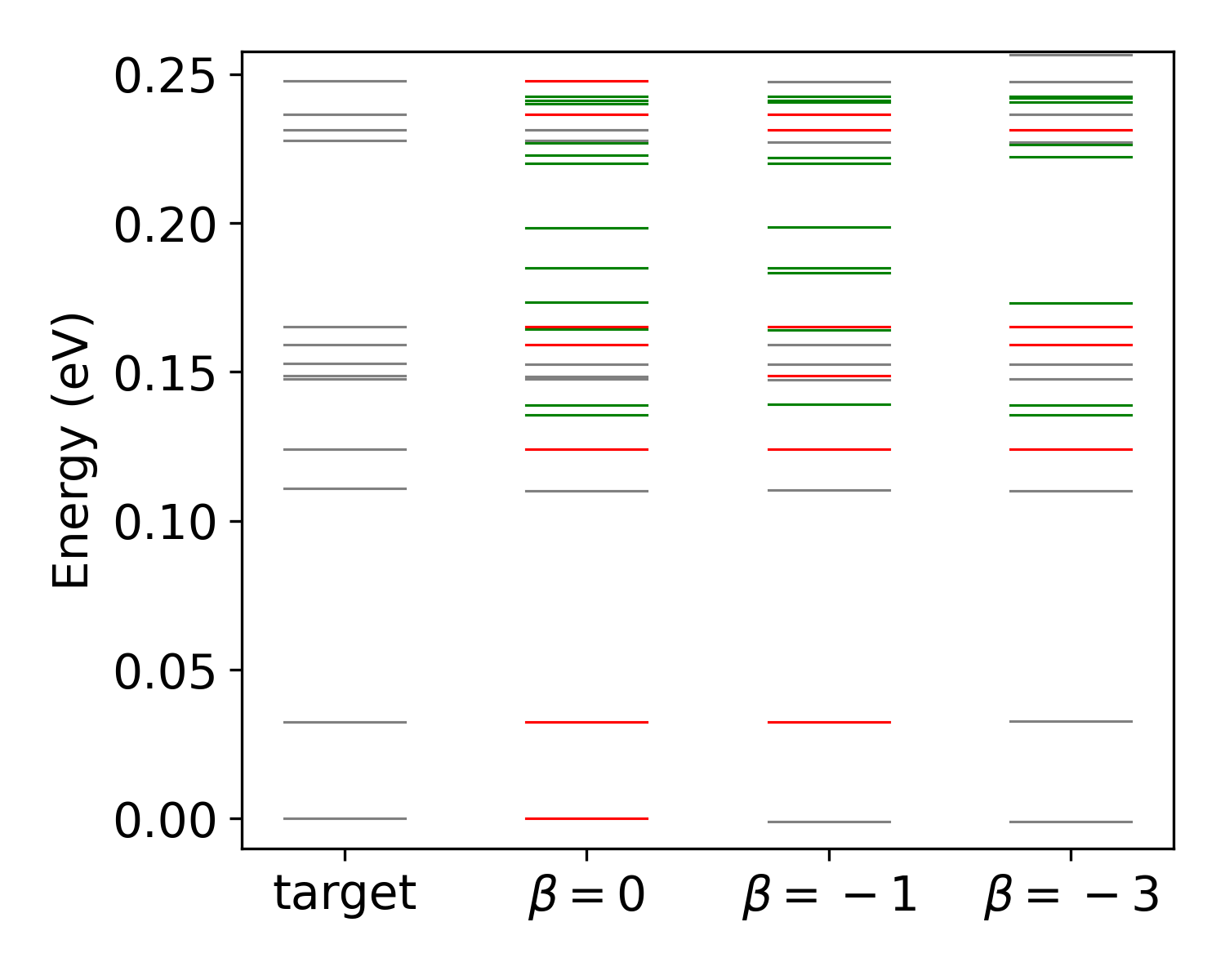}
        \put(-20, 75){\textbf{(c) WGG}}
    \end{overpic}
    \caption{The accumulative results for (a) WG (7d), (b) GGF (9d) and (c) WGG (9d).}
    \label{figure:resultsofbigmol}
\end{figure}

Figure \ref{figure:resultsofbigmol} shows the accumulative results for WG (7d), GGF (9d) and WGG (9d).

\section{Detailed results of real space search}

Algorithm S1 shows the real space search workflow.

\begin{algorithm}[htp]
\caption{Real space search workflow}
\begin{algorithmic}[1]
\Require $noise, M, k$
\State $\textbf{DataPool} = \emptyset$
\State $\textbf{StableComformers} = \emptyset$
\For {$i = 1 \ldots M$}
    \State $vector = \textbf{RamdomVector()}$
    \State $atoms = \textbf{Vec2Atoms}(vector)$
    \State $energy = \textbf{DFTEnergy}(atoms)$
    \State $\textbf{DataPool} \leftarrow \{vector, energy\}$
    \If {$(i\equiv 0\mod{k})$}
        \State $\textbf{Initialize}(\textbf{GP}, \textbf{DataPool})$
        \State $\textbf{Optimize}(\textbf{GP}, noise)$
        \For {$vector, energy \in \textbf{DataPool}$}
            \State $\textbf{Optimize}(vector, \textbf{GP})$
            \State $atoms = \textbf{Vec2Atoms}(vector)$
            \State $\textbf{Optimize}(atoms)$
            \State $\textbf{StableComformers} \leftarrow atoms$
        \EndFor
    \EndIf
\EndFor
\State \textbf{Return StableComformers}
\end{algorithmic}
\end{algorithm}

\begin{figure}[htp]
    \centering
    \includegraphics[width=0.8\textwidth]{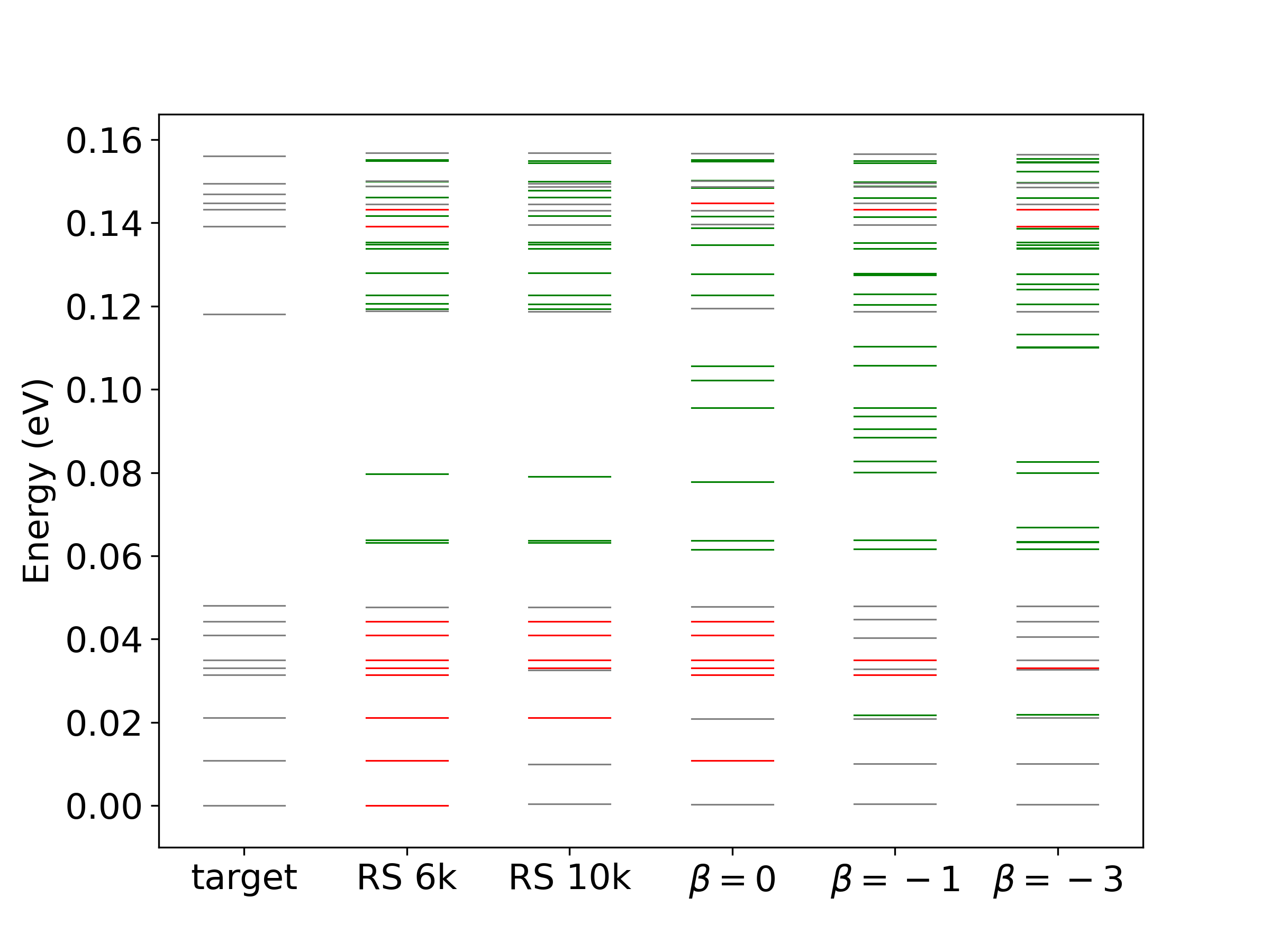}
    \caption{Comparison of LOLS and real space search workflow on GFA. RS 6k and RS 10k represent the results of the real space search workflow with 6000 and 10000 samples. }
    \label{figure:randomgp}
\end{figure}

Figure \ref{figure:randomgp} is the comparison of the GFA results from real space search workflow (Algorithm 2) and the LOLS. The  real space search workflow acquired 6000 and 10000 samples. LOLS acquired 6000 samples with different values of $\beta$. The real space search workflow missed the top eight lowest energy targets with 6000 samples and remained to perform poorly up to 10000 samples.

\section{Analysis of the final data pools of cysteine, WG, and GFA}

\begin{figure}[htp]
    \centering
    \includegraphics[width=0.8\textwidth]{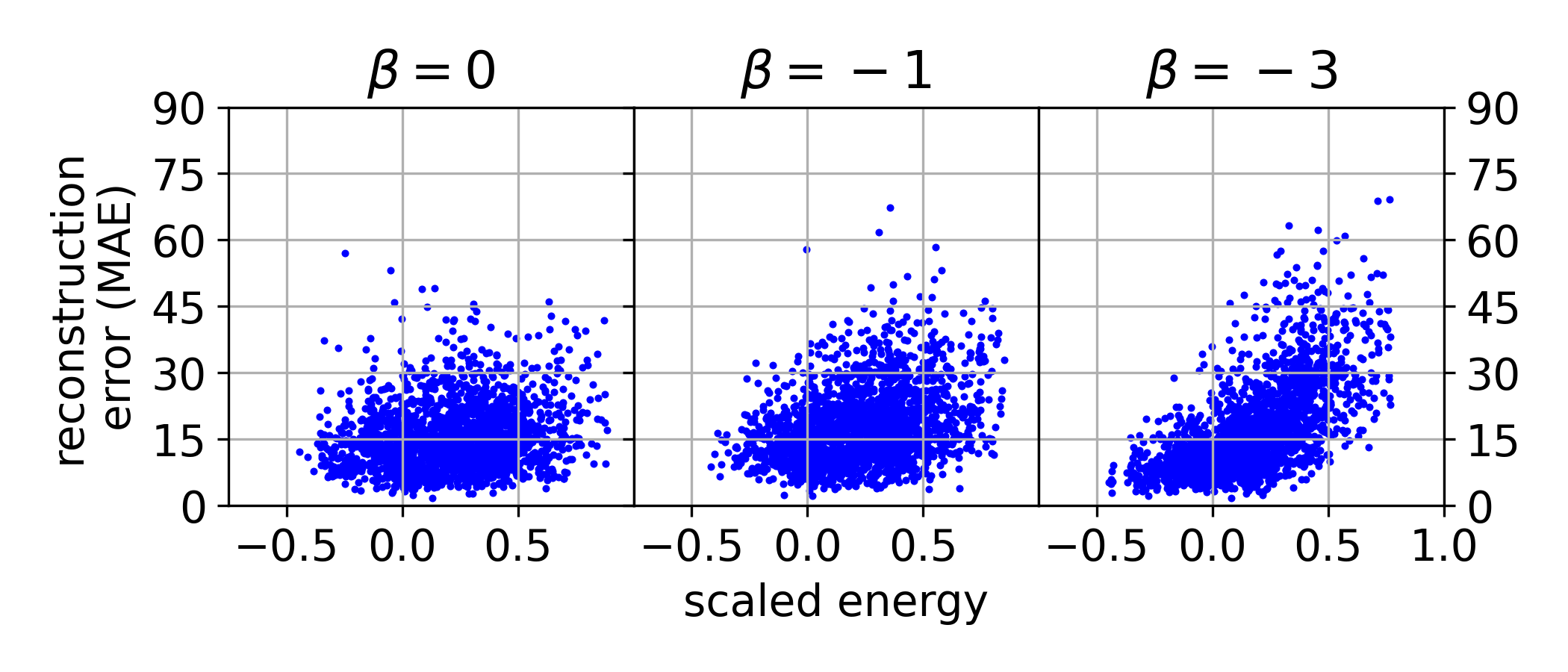}
    \caption{ The relationship between the reconstruction error (MAE of dihedral angles)
and the scaled energy in the last iteration of LOLS for cysteine.}
    \label{figure:maevsenergy}
\end{figure}

Figure \ref{figure:maevsenergy} gives the relationship between the reconstruction error (MAE of dihedral angles) and the scaled energy in the last iteration of LOLS for cysteine. A negative $\beta$ makes the data with lower energy have less reconstruction loss and causes the workflow to acquire more lower-energy data.

\begin{figure}[htp]
    \centering
    \includegraphics[width=0.8\textwidth]{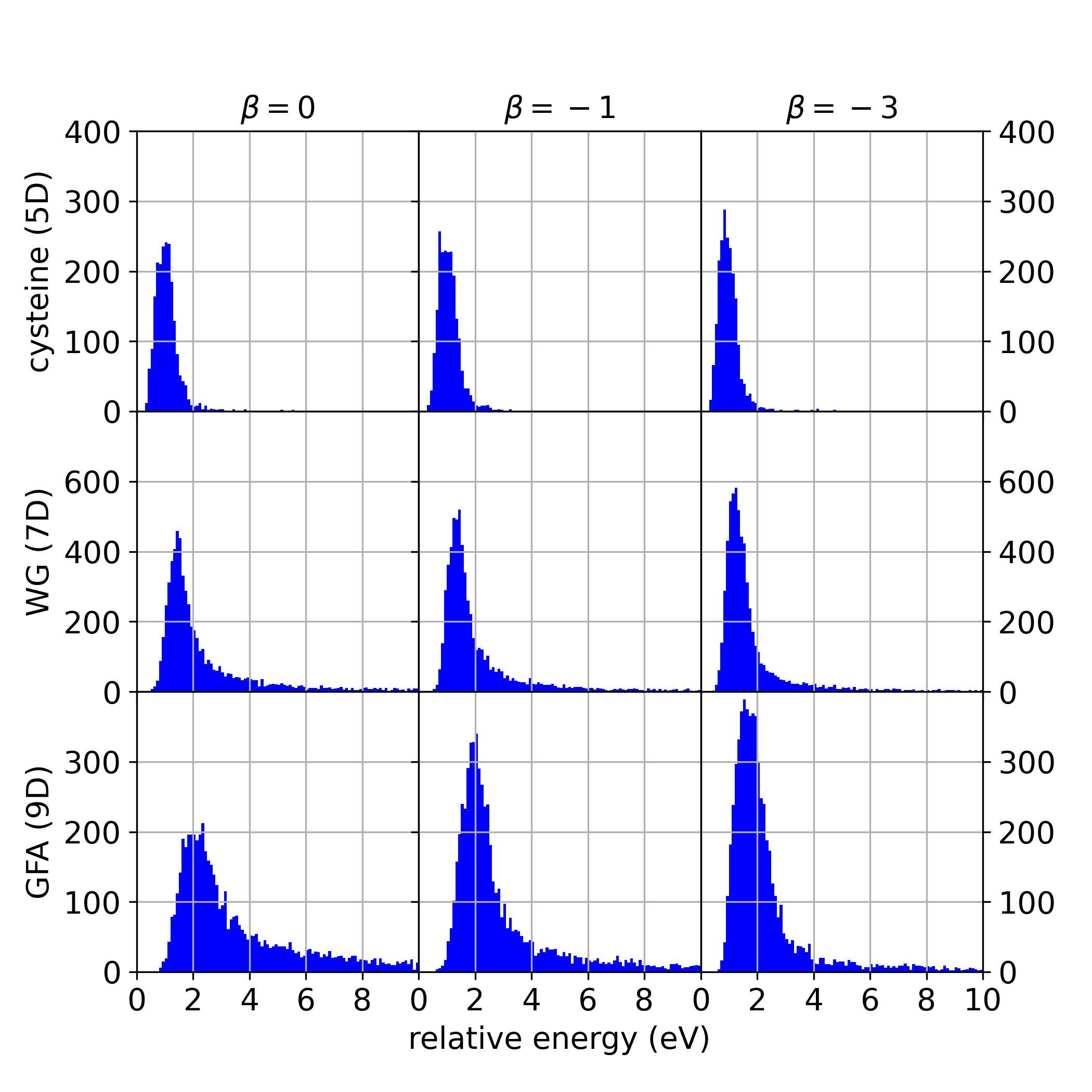}
    \caption{The energy distribution of the data in the last LOLS iteration for cysteine, WG and GFA.}
    \label{figure:energymodel}
\end{figure}

Figure \ref{figure:energymodel} shows the energy distribution of the data in the last LOLS iteration for cysteine, WG and GFA. The iteration number for Cysteine, WG, and GFA are 40, 120, and 120. The energy zero point is set to the global minima of the molecules. $\beta=-3$ makes LOLS obtain more lower energy samples. And the tendency becomes more significant as the dimension increases from five to nine. On the other hand, the ratio of higher energy ($> 4$ eV) increases as the dimension increases. The reason is that molecules with more atoms have a higher probability of collisions between atoms which causes high DFT energies
